\documentclass[11pt]{article}
\usepackage[utf8]{inputenc}
\usepackage[margin=1in]{geometry}
\usepackage[authoryear,round,sort]{natbib}
\usepackage{graphicx}
\usepackage{amsthm}
\usepackage{amsmath}
\usepackage{amssymb}
\usepackage{placeins}
\usepackage{textcomp,gensymb}
\usepackage{caption, subcaption}
\usepackage{comment}
\usepackage{booktabs}
\usepackage{adjustbox}
\usepackage{color}
\usepackage{tikz}
\usepackage{url}
\usepackage{multirow}
\usepackage{afterpage}
\usepackage{lscape}

\usepackage{array}
\newcolumntype{L}[1]{>{\raggedright\let\newline\\\arraybackslash\hspace{0pt}}m{#1}}
\newcolumntype{C}[1]{>{\centering\let\newline\\\arraybackslash\hspace{0pt}}m{#1}}
\newcolumntype{R}[1]{>{\raggedleft\let\newline\\\arraybackslash\hspace{0pt}}m{#1}}

\usepackage{setspace}
\usepackage{titlesec}

\titleformat*{\section}{\large\bfseries}
\titleformat*{\subsection}{\bfseries}
\titleformat*{\subsubsection}{\bfseries}
\titleformat*{\paragraph}{\itshape}
\titleformat*{\subparagraph}{\bfseries}

\title{A large-scale analysis of racial disparities\\in police stops across the United States\thanks{This work was supported by the John S. and James L. Knight Foundation, and by the Hellman Fellows Fund. 
EP acknowledges support from a Hertz Fellowship and an NDSEG Fellowship,
and SC acknowledges support from the Karr Family Graduate Fellowship.
All data and analysis code are available at \protect\url{https://openpolicing.stanford.edu}.
Correspondence may be addressed to Sharad Goel at scgoel@stanford.edu.
}}
\author{Emma Pierson\\Stanford University 
\and Camelia Simoiu\\Stanford University 
\and Jan Overgoor\\Stanford University 
\and Sam Corbett-Davies\\Stanford University
\and Vignesh Ramachandran\\Stanford University 
\and Cheryl Phillips\\Stanford University 
\and Sharad Goel\\Stanford University}
\date{}

\begin{document}

\singlespacing
\maketitle
\thispagestyle{empty}

\begin{abstract}
To assess racial disparities in police interactions with the public,
we compiled and analyzed a dataset  
detailing over 60 million 
state patrol stops conducted in 20 U.S. states between 2011 and 2015.
We find that black drivers are stopped more often than white drivers relative to their share of the driving-age population, but that Hispanic drivers are stopped less often than whites.
Among stopped drivers---and after controlling for
age, gender, time, and location---blacks and Hispanics are more
likely to be ticketed, searched, and arrested than white drivers.
These disparities may reflect differences in driving behavior,
and are not necessarily the result of bias.
In the case of search decisions, we explicitly test for discrimination
by examining both the rate at which drivers are searched 
and the likelihood searches turn up contraband.
We find evidence that the bar for searching 
black and Hispanic drivers is lower than for searching whites.
Finally, we find that legalizing recreational marijuana 
in Washington and Colorado reduced the total number of searches and 
misdemeanors for all race groups,
though a race gap still persists.
We conclude by offering recommendations for improving
data collection, analysis, and reporting by law enforcement agencies.
\end{abstract}

\newpage
\setcounter{page}{1}

\section{Introduction}
More than 20 million Americans are stopped each year for traffic violations,
making this one of the most common ways in which the public interacts with 
the police~\citep{langton2013}.
Due to a lack of comprehensive data,
it has been difficult to rigorously assess the manner and 
extent to which race plays a role in traffic stops~\citep{epp2014}.
The most widely cited national statistics come from the Police-Public Contact Survey (PPCS),
which is based on a nationally representative sample of approximately 50,000 people who report having been recently stopped by the police~\citep{ppcs}.
In addition to such survey data, some local and state agencies have released periodic reports on
traffic stops in their jurisdictions, and have also
made their data available to researchers for analysis ~\citep{antonovics2009,simoiu2016,anwar2006,ridgeway2009,ridgeway2006,ryan2016,rojek2004influence,smith2001,warren2006,hetey2016,vermont_study,voigt2017}.
While useful, these datasets provide only a partial picture.
For example, there is concern that 
the PPCS, like nearly all surveys, suffers from selection bias and recall errors.
Data released directly by police departments are potentially more complete,
but are available only for select agencies, are typically 
limited in what is reported, 
and are inconsistent across jurisdictions.

Here we analyze a unique dataset detailing more than 60 million 
state patrol stops conducted in 20 states between 2011 and 2015.
We compiled this dataset through a series of public records requests filed with all 50 states,
and we are redistributing these records in a standardized form to facilitate future analysis. 
Our statistical analysis of these records proceeds in three steps.
First, we quantify racial disparities in stop rates and post-stop outcomes.
Adjusting for age, gender, location and year, we find that black drivers are stopped more often than white drivers relative to their share of the driving-age population, but find that Hispanic drivers
are stopped less often than whites.
After being stopped, black and Hispanic drivers
are more likely than whites to be ticketed, searched, and arrested.
Such disparities may stem from a combination of factors---including differences in driving behavior---and are not necessarily the result of racial bias.
In the second phase of our analysis, 
we investigate the degree to which these differences may result from discrimination, focusing on search decisions.
By examining both the rate at which searches occur
and the success rate of these searches,
we find evidence that the bar for searching black and Hispanic drivers
is lower than for searching white drivers.
Finally, we examine the effects of drug policy on stop outcomes.
We find that legalizing recreational marijuana in Washington and
Colorado reduced both search and misdemeanor rates for white, black, and Hispanic drivers, though a relative gap persists.
We conclude by suggesting best-practices for
data collection, analysis, and reporting 
by law enforcement agencies.

\section{Compiling a national database of traffic stops}

%
%
\begin{figure}
\centering
\includegraphics[width=.6\paperwidth]{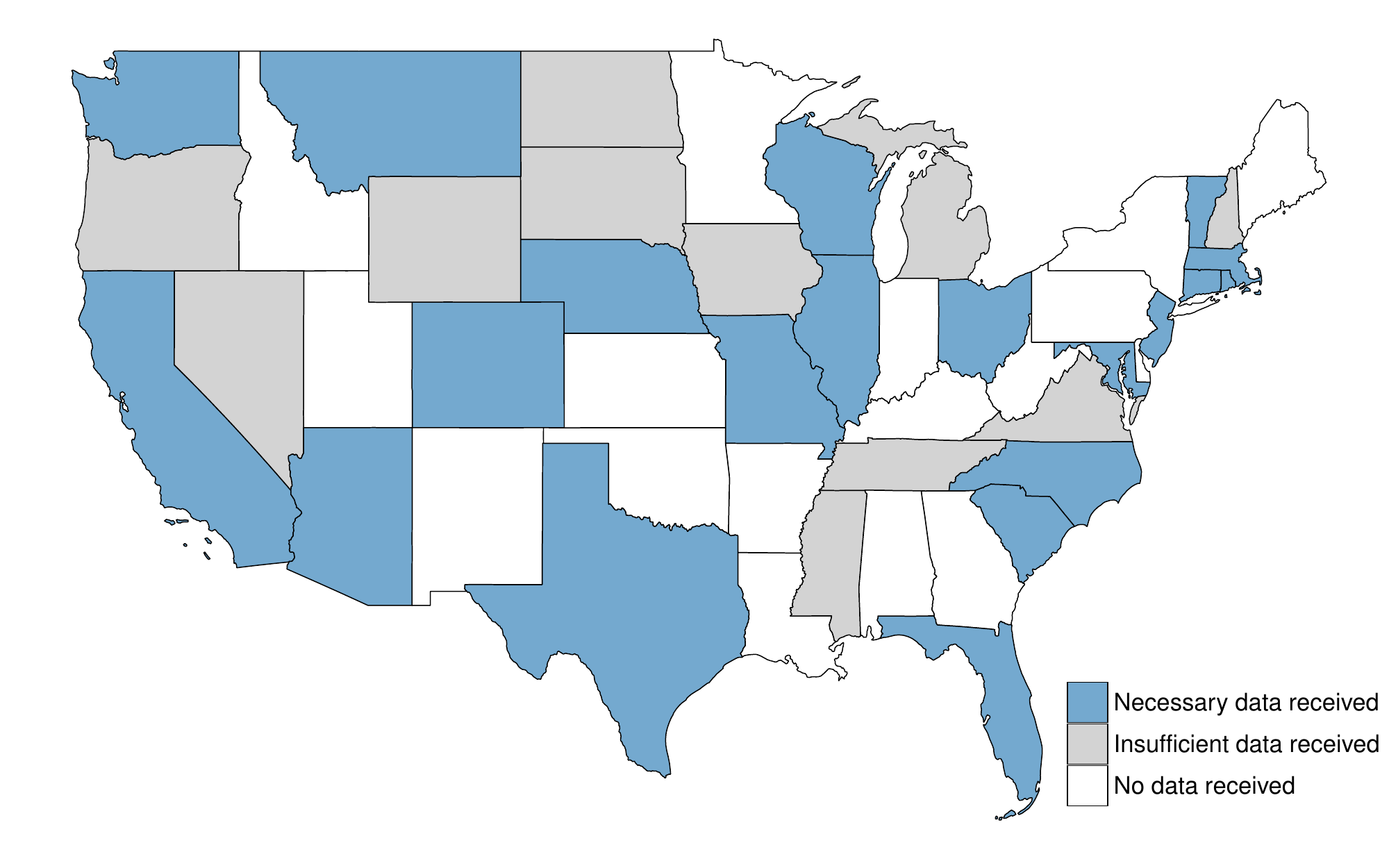}
\caption{
\emph{We collected detailed information on over 60 million state patrol stops conducted in 20 states between 2011 and 2015.
An additional 11 states provided data that are
insufficient to assess racial disparities,
and 19 states have not provided any data (including Hawaii and Alaska).}}
\label{figure:map}
\end{figure}

\subsection{Data collection}

To assemble a national dataset of traffic stops, 
we first identified which state law enforcement agencies electronically maintain traffic stop records that, at a minimum, include the race of the stopped driver.
Of the 50 state agencies,
7 did not respond to our request for information or did not disclose whether any data were collected;
an additional 9 agencies do not compile stop records electronically or reported that they were unable to send their data to us in electronic form;
and 3 state agencies keep electronic records but do not track the race of stopped drivers
(see Table~\ref{tab:states_status} for details).
For the remaining 31 states, we filed public records requests for detailed information on each stop conducted since 2005.

To date, we have collected data on approximately 136 million state patrol stops in  31 states.
Of these, we exclude 11 states from our analysis because the obtained data
were insufficient to assess racial disparities (e.g., the race of the stopped driver was not regularly recorded, or only a non-representative subset of stops was provided).
In the remaining 20 states that we consider, 18 provided data for each individual stop. In the other two---Missouri and Nebraska---only summary data were provided, but these summaries were sufficiently granular to allow for statistical analysis. 
For consistency in our analysis, we restrict to stops occurring in 2011--2015, as many states did not provide data on earlier stops.
We also limit our analysis to drivers classified as white, black or Hispanic, as there are relatively few recorded stops of drivers in other race groups.
Our primary dataset thus consists of 63.7 million state patrol stops from 20 states (Figure~\ref{figure:map}).

\subsection{Data normalization}

Each state provided the stop data in idiosyncratic formats with varying levels of specificity, and so we used a variety of automated and manual procedures to create
the final dataset.
For each recorded stop, we attempted to extract and normalize
the date and time of the stop;
the county or state patrol district in which the stop took place;
the race, gender and age of the driver; 
the stop reason;
whether a search was conducted;
the legal justification for the search (e.g., ``probable cause'' or ``consent'');
whether contraband was found during a search;
and the stop outcome (e.g., a citation or an arrest).
We describe our procedures for normalizing each of these covariates in the Appendix.
As indicated in Table~\ref{table:summary}, 
the availability of information varies significantly across states.
We therefore restrict each of our specific analyses 
to the corresponding subset of states for which we have the required fields.

In many states, more than one row in the raw data appeared to refer to the same stop. 
For example, in several states each row referred to one \emph{violation}, not one stop. 
We detected and reconciled such duplicates by inspecting columns with granular values. 
For example, in Colorado we counted two rows as duplicates if they had the same officer identification code, officer first and last name, driver first and last name, driver birth date, stop location (precise to the milepost marker), and stop date and time (precise to the minute). 

%
%
\newcommand*{\thead}[1]{\multicolumn{1}{c}{#1}}

\begin{table}[t!]
\tiny
\centerline{
\begin{tabular}{clrlccccccccccccc}
 &  \multirow{2}{*}{State}  & \thead{\multirow{2}{*}{Stops}} & \thead{Time}  & \thead{Stop} & \thead{Stop} & \thead{Stop} & \thead{Driver} & \thead{Driver} & \thead{Stop} & \thead{Search} & \thead{Search} & \thead{Contraband} & \thead{Stop}  \\
 &  &  & \thead{Range} & \thead{Date} & \thead{Time} & \thead{Location} & \thead{Gender} & \thead{Age} & \thead{Reason} & \thead{Conducted} & \thead{Type} & \thead{Found} & \thead{Outcome}  \\
\hline
1 & Arizona &  2,039,781 & 2011-2015 & $\bullet$ & $\bullet$ & $\bullet$ & $\bullet$ &  &  & $\bullet$ &  & $\bullet$ & $\bullet$ \\ 
  2 & California & 19,012,414 & 2011-2015 & $\bullet$ &  & $\bullet$ & $\bullet$ &  & $\bullet$ & $\bullet$ & $\bullet$ &  & $\bullet$ \\ 
  3 & Colorado &  1,674,619 & 2011-2015 & $\bullet$ & $\bullet$ & $\bullet$ & $\bullet$ & $\bullet$ & $\bullet$ & $\bullet$ & $\bullet$ & $\bullet$ &  \\ 
  4 & Connecticut &    310,969 & 2013-2015 & $\bullet$ & $\bullet$ & $\bullet$ & $\bullet$ & $\bullet$ & $\bullet$ & $\bullet$ & $\bullet$ & $\bullet$ & $\bullet$ \\ 
  5 & Florida &  4,002,547 & 2011-2015 & $\bullet$ & $\bullet$ & $\bullet$ & $\bullet$ & $\bullet$ & $\bullet$ & $\bullet$ & $\bullet$ &  & $\bullet$ \\ 
  6 & Illinois &  1,528,340 & 2011-2015 & $\bullet$ & $\bullet$ & $\bullet$ & $\bullet$ & $\bullet$ & $\bullet$ & $\bullet$ &  & $\bullet$ & $\bullet$ \\ 
  7 & Maryland &    578,613 & 2011-2014 &  &  &  & $\bullet$ &  & $\bullet$ & $\bullet$ & $\bullet$ & $\bullet$ & $\bullet$ \\ 
  8 & Massachusetts &  1,773,546 & 2011-2015 & $\bullet$ &  & $\bullet$ & $\bullet$ & $\bullet$ &  & $\bullet$ & $\bullet$ & $\bullet$ & $\bullet$ \\ 
  9 & Missouri &  1,906,797 & 2011-2015 &  &  &  &  &  &  & $\bullet$ &  & $\bullet$ &  \\ 
  10 & Montana &    547,115 & 2011-2015 & $\bullet$ & $\bullet$ & $\bullet$ & $\bullet$ & $\bullet$ & $\bullet$ & $\bullet$ & $\bullet$ &  & $\bullet$ \\ 
  11 & Nebraska &    840,764 & 2011-2014 &  &  &  &  &  &  & $\bullet$ &  &  &  \\ 
  12 & New Jersey &  2,069,123 & 2011-2015 & $\bullet$ & $\bullet$ & $\bullet$ & $\bullet$ &  & $\bullet$ &  &  &  & $\bullet$ \\ 
  13 & North Carolina &  3,500,180 & 2011-2015 & $\bullet$ &  & $\bullet$ & $\bullet$ & $\bullet$ & $\bullet$ & $\bullet$ & $\bullet$ & $\bullet$ & $\bullet$ \\ 
  14 & Ohio &  4,660,935 & 2011-2015 & $\bullet$ & $\bullet$ & $\bullet$ & $\bullet$ &  &  & $\bullet$ &  &  &  \\ 
  15 & Rhode Island &    229,691 & 2011-2015 & $\bullet$ & $\bullet$ & $\bullet$ & $\bullet$ & $\bullet$ & $\bullet$ & $\bullet$ & $\bullet$ & $\bullet$ & $\bullet$ \\ 
  16 & South Carolina &  3,696,801 & 2011-2015 & $\bullet$ &  & $\bullet$ & $\bullet$ & $\bullet$ &  & $\bullet$ &  & $\bullet$ & $\bullet$ \\ 
  17 & Texas & 10,239,721 & 2011-2015 & $\bullet$ & $\bullet$ & $\bullet$ & $\bullet$ &  & $\bullet$ & $\bullet$ & $\bullet$ & $\bullet$ & $\bullet$ \\ 
  18 & Vermont &    250,949 & 2011-2015 & $\bullet$ & $\bullet$ & $\bullet$ & $\bullet$ & $\bullet$ & $\bullet$ & $\bullet$ & $\bullet$ & $\bullet$ & $\bullet$ \\ 
  19 & Washington &  4,053,099 & 2011-2015 & $\bullet$ & $\bullet$ & $\bullet$ & $\bullet$ & $\bullet$ & $\bullet$ & $\bullet$ & $\bullet$ & $\bullet$ & $\bullet$ \\ 
  20 & Wisconsin &    827,028 & 2011-2015 & $\bullet$ & $\bullet$ & $\bullet$ & $\bullet$ &  & $\bullet$ & $\bullet$ & $\bullet$ & $\bullet$ & $\bullet$ \\ 
   & \textbf{Total} & \textbf{63,743,032} &  &  &  &  &  &  &  &  &  &  &  \\ 
\end{tabular}
}
\vspace{2mm}
\caption{\emph{Availability of data in the 20 states comprising our primary analysis, 
where for each column a solid circle signifies 
data are available for at least 70\% of the stops. 
For all states except Illinois, North Carolina, and Rhode Island, ``stop location'' refers to county; for these three states, it refers to a similarly granular location variable, as described above.
}}
\label{table:summary}
\end{table}

\subsection{Error correction}

The raw data in many states contain errors. We ran numerous automated checks to detect and correct these where possible, although some errors likely remain
due to the complex nature of the data. 
For example, after examining the distribution of recorded values in each state, we discovered a spurious density of stops in North Carolina listed as occurring at precisely midnight.
As the value ``00:00'' was likely used to indicate missing information, we treated it as such. 

Past work suggests that Texas state patrol officers 
incorrectly recorded many Hispanic drivers as white.\footnote{See: \url{http://kxan.com/investigative-story/texas-troopers-ticketing-hispanics-motorists-as-white/}} 
To investigate and correct for this issue,
we impute Hispanic ethnicity from surnames in the
three states for which we have name data: 
Texas, Arizona, and Colorado.
To do so, we use a dataset from the U.S. Census Bureau that estimates the racial and ethnic distribution of people with a given surname, for surnames occurring at least 100 times \citep{word2008demographic}.\footnote{\url{http://www.census.gov/topics/population/genealogy/data/2000_surnames.html}} 
To increase the matching rate, we perform minor string edits to the names, including removing punctuation and suffixes (e.g., ``Jr.'' and ``II''), and consider only the longest word in multi-part surnames.
Following past work~\citep{word1996building,2009ortega},
we define a name as ``Hispanic-affiliated'' if at least 75\% of
people with that name identify as Hispanic, according to the 2000 Census; 
we note that 90\% of those with Hispanic-affiliated names identify as Hispanic.
Among drivers with Hispanic-affiliated names, 
the proportion labeled as Hispanic in the raw data is considerably lower in Texas (37\%) than in either Arizona (79\%) or Colorado (70\%),
corroborating past results.
Though imperfect, we re-categorize as ``Hispanic'' all drivers 
in Texas with Hispanic-affiliated names who were
originally labeled ``white'' or had missing race data.

Our complete data cleaning pipeline is extensive, 
requiring subjective decisions and thousands of lines of code. 
For transparency and reproducibility, we have released the raw data, 
the standardized data, and code to clean and analyze the records 
at \url{https://openpolicing.stanford.edu}.

\section{Stop rates and post-stop outcomes}

We begin our analysis by examining the extent to which
there are racial disparities in stop, citation, search, and arrest rates.
The disparities we discuss below likely 
result from a combination of complex factors, 
and do not necessarily reflect racial bias.
Regardless of the mechanism, however, we quantify these disparities
in order to better understand the differential impact 
policing has on minority communities.

\begin{figure}[t]
\centering
\includegraphics[height=5cm]{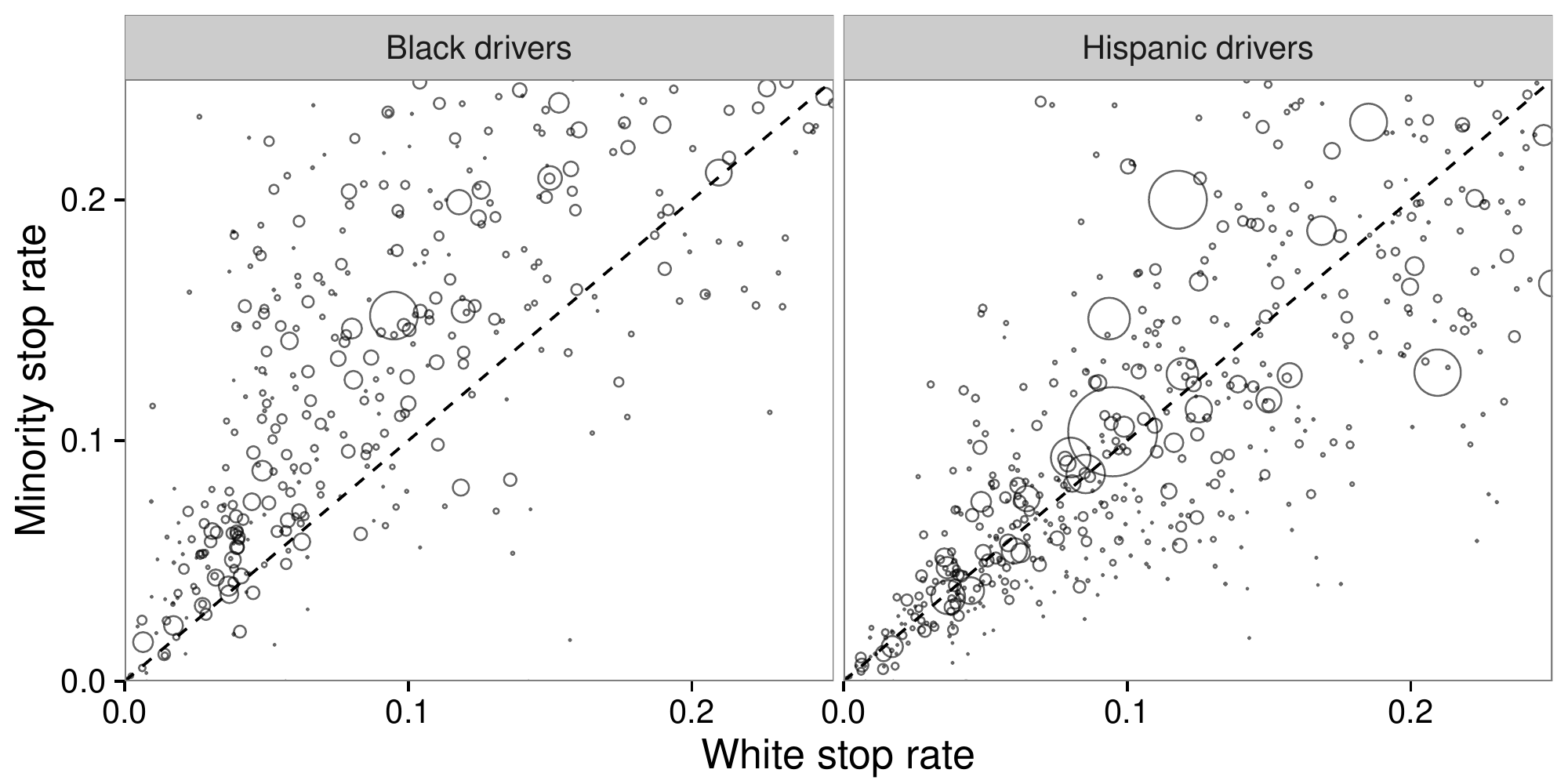}
\caption{\emph{Stops per person of driving age, stratified by race and location, where points are sized proportional to the number of stops. The plots cover 16 states for which we have location data. Within location, black drivers are often stopped more often than white drivers;
Hispanic drivers are generally stopped at similar rates as whites.
}}
\label{figure:benchmark}
\end{figure}

\subsection{Stop rates}

We first estimate the rate at which white, black, and Hispanic drivers are stopped,
relative to their share of the driving-age population~\citep{smith2001}.
Although there are a variety of benchmarks one might consider~\citep{new_jersey_state_patrol_report,alpert2004,engel2004},
the driving-age population has the unique distinction 
of being readily available in nearly every jurisdiction,
and it is accurately estimated by the U.S. Census Bureau;\footnote{We use the intercensal estimates produced by the U.S. Census Bureau, available at \url{https://www2.census.gov/programs-surveys/popest/datasets/2010-2015/counties/asrh/cc-est2015-alldata.csv} or from our Open Policing website.}
we note, however, that this benchmark does not account for possible race-specific differences in driving behavior, including amount of time spent on the road and adherence to traffic laws.

Figure~\ref{figure:benchmark} shows stop rates of black and Hispanic drivers relative to whites, disaggregated by location. 
Each point in the plot corresponds to either the county or similar geographic unit
in which the stop was made.
We find that Hispanics are stopped at similar rates as whites in most jurisdictions;
black drivers, however, are stopped more often than whites in over 80\% of the locations we consider. 

We next estimate race-specific stop rates after adjusting for 
driver demographics (age and gender), stop location, and stop year;
age was binned into the categories 16--19, 20--29, 30--39, 40--49, and 50+ years-old.
In our primary analysis, we fit a negative binomial regression, where we benchmark to the 
census-estimated driving-age population:
\begin{equation*}
y_{rag\ell y} \sim 
\text{NegBin}\left( n_{rag\ell y}e^{\mu + \alpha_r + \beta_a + \gamma_g + \delta_{\ell} + \epsilon_y}, \phi\right)
\end{equation*}
where $y_{rag\ell y}$ is the observed number of stops in a group defined by
race, age, gender, location, and year, $n_{rag\ell y}$ is the corresponding census benchmark,
and $\alpha_r$ are the key race coefficients (we set $\alpha_{\text{white}} = 0$). 
The negative binomial distribution is parameterized such that if $Y \sim \text{NegBin}(\mu, \phi)$, then 
$\mathbb{E}[Y] = \mu$ and $\text{Var}[Y] = \mu + \mu^2/\phi$.
The parameter $\phi$ allows for overdispersion, and is estimated from the data. 

Table~\ref{tab:demographic_regression_coefs} (first column) shows the estimated race, gender and age coefficients; we further estimate $\hat{\phi} = 3.9$. 
After controlling for gender, age, location, and year,
we find that blacks are stopped
at 1.4 times the rate at which whites are stopped ($e^{0.37} = 1.4$),
and Hispanics are stopped at 0.7 times the white stop rate ($e^{-0.40} = 0.7$).
To help interpret these numbers, Table~\ref{tab:regressions}
shows stop rates for a typical 20-29 year-old male driver:
the per-capita stop rate is 0.42 for blacks, 0.29 for whites, and 0.19 for Hispanics.

As shown in Figure~\ref{figure:benchmark}, Hispanic drivers are stopped at similar rates as whites when controlling only for location. But Hispanic drivers are more likely to be young, and young drivers are more likely to be stopped. 
As a result, after additionally adjusting for age (and other covariates) in the regression above, 
we find Hispanics are stopped at a lower rate than whites. 
This lower estimated rate is 
consistent with self-reports collected as part of the PPCS~\citep{ppcs}. 
With the PPCS data, we used logistic regression to estimate the likelihood a 
respondent would report having been stopped by the police while driving,
where we controlled for the respondent's race, age, gender, and size of city.
We found that Hispanic respondents were less likely than white drivers to report having been stopped (odds ratio = 0.85). 
This result is in line with a similar analysis of the same PPCS data~\citep{medina2016accounting}.

To check the robustness of the observed racial disparities, we additionally fit stop rate regressions using a Poisson model with sandwich errors, and using a quasi-Poisson model~\citep{ver2007quasi,gardner1995regression}.
We report these results in Table~\ref{tab:all_regression_coefficients} (first three rows). 
The signs of the race coefficients are the same under all three specifications,
but the estimated effect sizes are somewhat larger in the negative binomial model
than in the two Poisson models (both of which necessarily yield identical coefficients).
We note that it is common for Poisson and negative binomial formulations to produce
somewhat different effect sizes~\citep{ver2007quasi}.

\begin{table}[t]
\centering
\begin{tabular}{rccccc}
 & Stop & Citation & Search & Consent search & Arrest \\ 
 \hline
  Black & 0.37 (0.01) & 0.18 (0.00) & 0.73 (0.01) & 0.77 (0.03) & 0.65 (0.01) \\ 
  Hispanic & -0.40 (0.01) & 0.29 (0.00) & 0.54 (0.01) & 0.62 (0.02) & 0.69 (0.01) \\ 
  Male & 0.72 (0.00) & 0.08 (0.00) & 0.58 (0.01) & 0.86 (0.02) & 0.43 (0.01) \\ 
  Age 20-29 & 0.65 (0.01) & -0.13 (0.01) & 0.13 (0.01) & -0.38 (0.03) & 0.38 (0.01) \\ 
  Age 30-39 & 0.47 (0.01) & -0.35 (0.01) & -0.06 (0.01) & -0.79 (0.03) & 0.30 (0.01) \\ 
  Age 40-49 & 0.25 (0.01) & -0.47 (0.01) & -0.37 (0.01) & -1.20 (0.04) & -0.04 (0.01) \\ 
  Age 50+ & -0.53 (0.01) & -0.68 (0.01) & -0.80 (0.01) & -1.82 (0.04) & -0.47 (0.01) \\ 
\end{tabular}
\caption{\emph{Coefficients and standard errors for stop rate and post-stop outcome models.}
\label{tab:demographic_regression_coefs}}
\end{table}

\begin{table}[t]
\centering
\begin{tabular}{lrrr}
& White & Black & Hispanic \\ 
  \hline
Stop rate & 0.29 & 0.42 & 0.19 \\ 
  Speeding citation & 72\% & 75\% & 77\% \\ 
  Search & 1.3\% & 2.7\% & 2.3\% \\ 
  Consent search & 0.1\% & 0.3\% & 0.3\% \\ 
  Arrest & 2.8\% & 5.3\% & 5.5\% \\ 
\end{tabular}
\vspace{2mm}
\caption{\emph{Model-estimated rates for a typical 20-29 year-old male. 
The ``speeding citation'' outcome corresponds to receiving a citation rather than a warning (or no penalty) when pulled over for speeding.
Negative binomial regression is used for stop rate (first row), benchmarked to the driving-age population; logistic regression is used for all other analyses. The stop rate regression includes controls for age, gender, stop location, and stop year; all other regressions additionally include controls for stop quarter, weekday, and hour (binned into three-hour segments).
}}
\label{tab:regressions}
\end{table}

\subsection{Citation, search, and arrest rates}

Stop rates are a natural starting point but are inherently 
difficult to interpret, in part because
results can be sensitive 
to the benchmark used. (We note that there are 
no readily available alternatives to the driving-age population.)
We thus now consider post-stop outcomes, 
starting with the rates at which white and 
minority drivers receive citations rather than warnings when pulled over for speeding. 

We use logistic regression to estimate racial disparities in the probability a  driver stopped for speeding is given a citation as opposed to a warning (or no penalty at all).  
In addition to driver age and gender, location, 
and year, we control for stop quarter, stop weekday, and stop hour, 
binned into eight 3-hour segments. 
(In the case of stop rates, we used negative binomial and Poisson models since we were estimating total counts; we could not control for time in that case because we lacked time-specific population benchmarks.)
Table~\ref{tab:demographic_regression_coefs} (second column) shows the 
estimated race, gender and age coefficients.
We find that black drivers have 19\% higher odds of receiving a citation than white drivers, 
and Hispanics have 34\% higher odds than whites. 
For typical young male drivers, Table~\ref{tab:regressions}
shows that 72\% of whites stopped for speeding receive a citation,
compared to 75\% and 77\% for black and Hispanic drivers, respectively.

\begin{figure}[t]
\centering
\includegraphics[height=5cm]{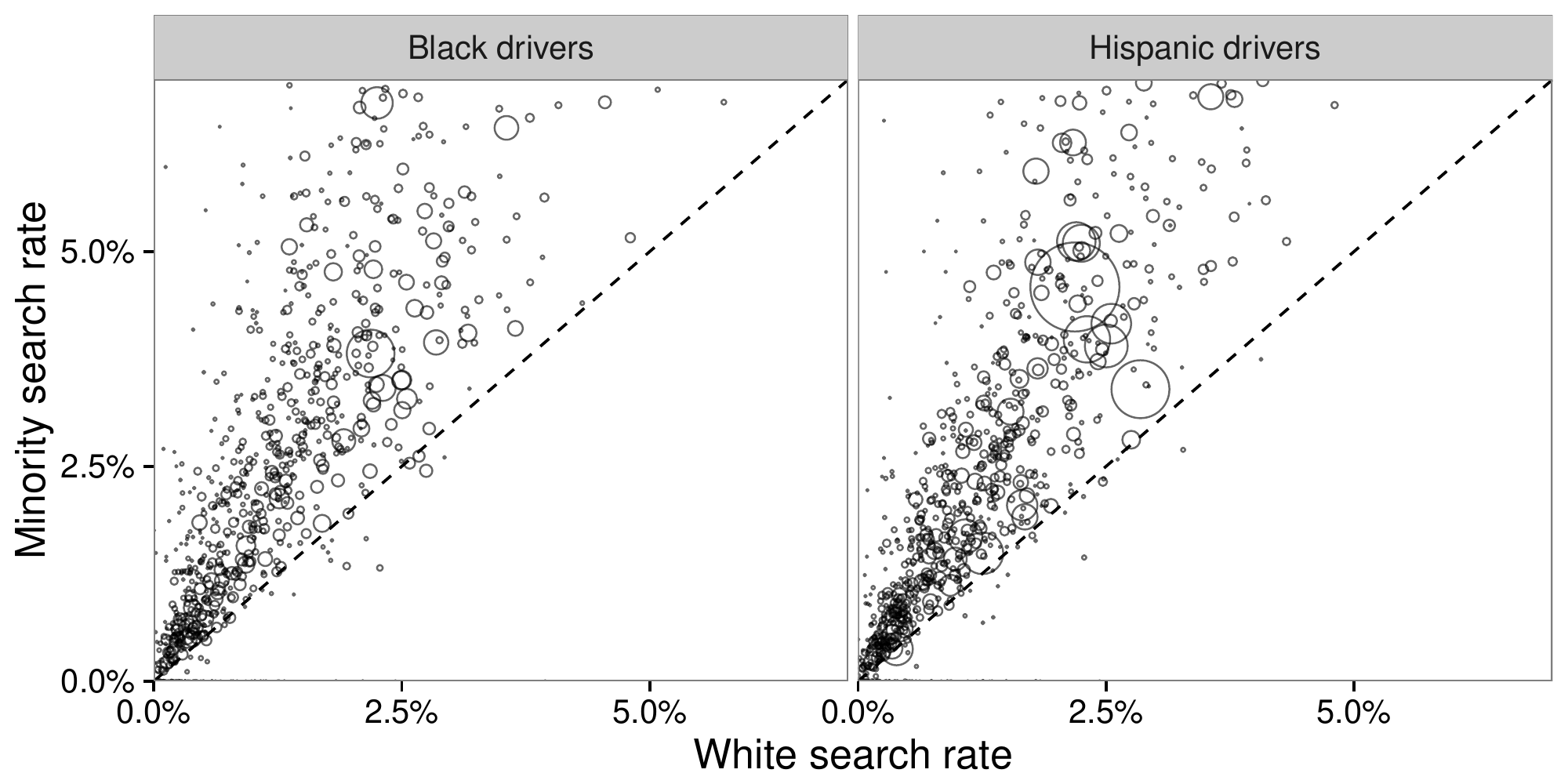}
\includegraphics[height=5cm]{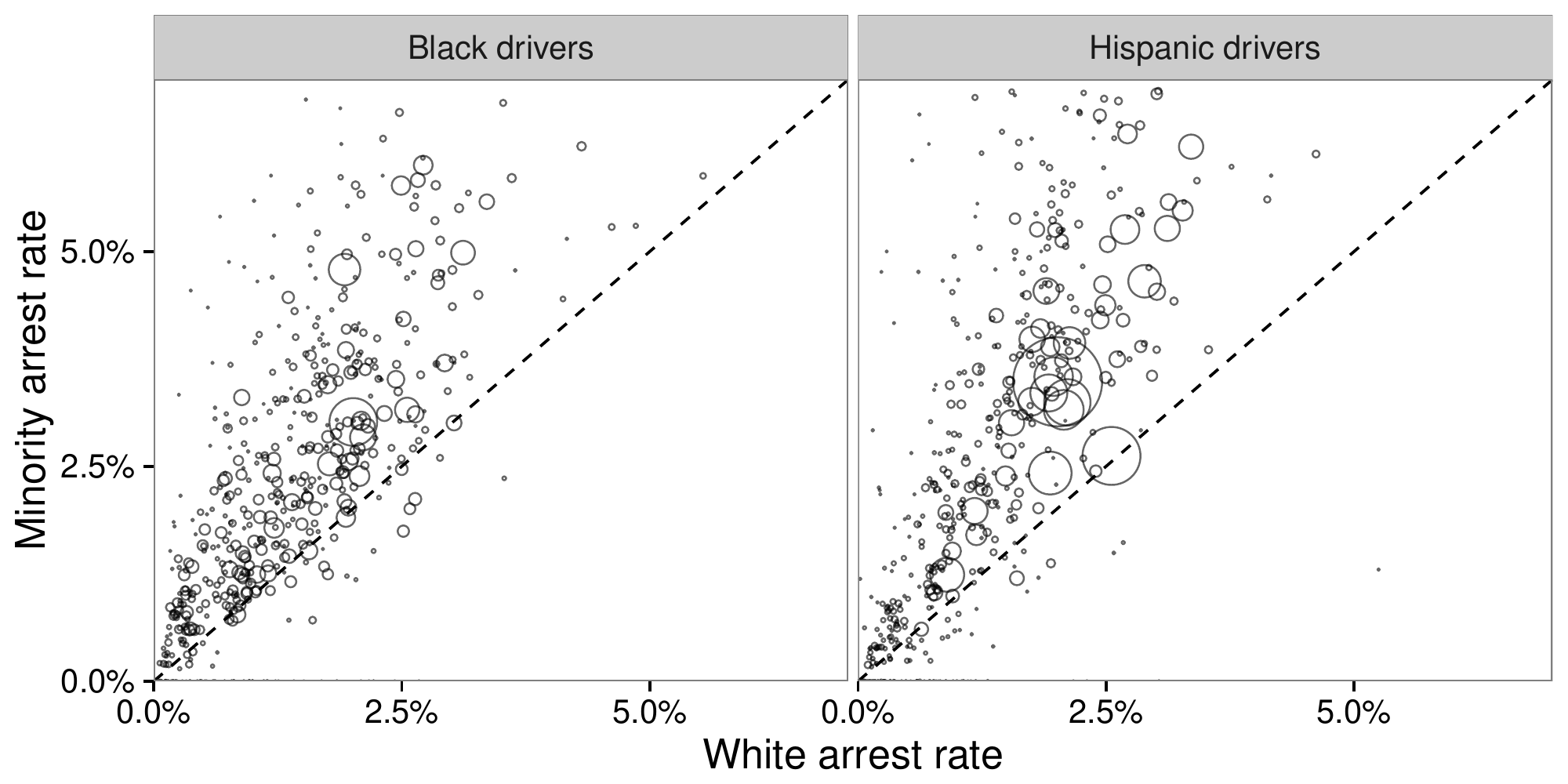}
\caption{\emph{Search rates (top) and arrest rates (bottom) by race and location among stopped drivers. In nearly every area, minorities are searched and arrested more often than whites. The search data cover 16 states, comprising a total of 56 million stops, and the arrest data include 40 million stops in 13 states.}}
\label{figure:search_rate}
\end{figure}

Next, we examine search rates.
After stopping a driver, officers may search both driver and vehicle for drugs, weapons, and other contraband when they suspect more serious criminal activity.
Aggregating across all states for which we have search data, white drivers are searched in 2.0\% of stops, compared to 3.5\% of stops for black motorists and 3.8\% for Hispanic motorists. 
Across jurisdiction, Figure~\ref{figure:search_rate} (top row)
shows that black and Hispanic motorists are consistently
searched at higher rates than white drivers.
After controlling for stop location, date and time, and driver age and gender---via logistic regression, as above---we
find that black and Hispanic drivers have approximately twice the
odds of being searched relative to white drivers (2.1 and 1.7, respectively, as shown in
Table~\ref{tab:demographic_regression_coefs}).

We now consider the subset of searches
conducted with consent, where officers must seek permission from drivers to search their vehicles. (In contrast, \emph{probable cause} searches do not require consent, but legally demand a high standard of evidence.\footnote{Officers may also conduct \emph{protective frisks} to search for weapons, a type of search that legally requires only \emph{reasonable suspicion}, a lower standard of evidence than probable cause. In our dataset, protective frisks occur much less frequently than probable cause and consent searches.})
We find that minority drivers are more likely than whites to undergo consent searches in the seven states for which we have reliable data 
(Colorado, Florida, Massachusetts, Maryland, North Carolina, Texas, and Washington); controlling for stop location, date and time, and driver age and gender,
we find that black drivers have 2.2 times the odds of whites and Hispanic drivers have 1.9 times the odds of whites of undergoing a consent search
(Table~\ref{tab:demographic_regression_coefs}).

Finally, we examine arrest rates.
In aggregate, black drivers are arrested in 2.8\% of stops and Hispanic drivers in 3.4\% of stops,
compared to 1.7\% for white drivers.  
Again controlling for driver age and gender, stop date and time, and stop location, we find that black drivers have 1.9 times the odds of being arrested, and Hispanic drivers have 2.0 times the odds of being arrested compared to white drivers (Figure~\ref{figure:search_rate} and Table~\ref{tab:demographic_regression_coefs}).

To assess the robustness of our results on citation, search, and arrest rates, 
we fit logistic regression models with five 
different sets of control variables, 
as described in Table \ref{tab:all_regression_coefficients}:
(1) driver race only; 
(2) driver race and county; 
(3) driver race, age, gender, and county;
(4) driver race, county, and stop time; and
(5) driver race, age, gender, county, and stop time.  
These five models were fit on the largest set of 
states for which the relevant information was available.
In nearly every case, the estimated race coefficients 
were positive and significant, indicating that black and Hispanic 
drivers were cited, searched, and arrested more often than white drivers.
There was one exception: we found a negative coefficient (-0.11) for Hispanic drivers
when estimating the likelihood of receiving a speeding citation when controlling only for race. This outlier occurs because Texas has an especially high fraction of Hispanic drivers and an especially low rate of citations. 
Finally, we confirmed that these racial disparities persist when we
alter the set of stops analyzed:
we find qualitatively similar results 
when we fit our models only on speeding stops; when we eliminate searches incident to arrest; and when we fit models on each state separately. 

\section{Testing for bias in search decisions} 

When stopped, 
black and Hispanic drivers are more likely to be 
issued citations, more likely to be searched,
and more likely to be arrested.
These disparities, however, are not necessarily the product
of discrimination. 
Minority drivers might, for example, carry contraband at higher rates than 
whites, and so elevated search rates may 
result from routine police work.
We now investigate whether bias plays a role in search decisions, a class of actions amenable to statistical analysis.

\subsection{The outcome test}

To start, we apply the \emph{outcome test}, 
originally proposed by \citet{becker1957, becker1993} to circumvent omitted variable bias in traditional tests of discrimination.
The outcome test is based not on the search rate, 
but on the \emph{hit rate}: the proportion of searches that successfully turn up contraband.
Becker argued that even if minority drivers are more likely to carry contraband, 
absent discrimination searched minorities should still be found to have contraband at the same rate as searched whites.
If searches of minorities are less often successful than searches of whites, 
it suggests that officers are applying a double standard,
searching minorities on the basis of less evidence.

\begin{figure}[t]
\centering
\includegraphics[height=5cm]{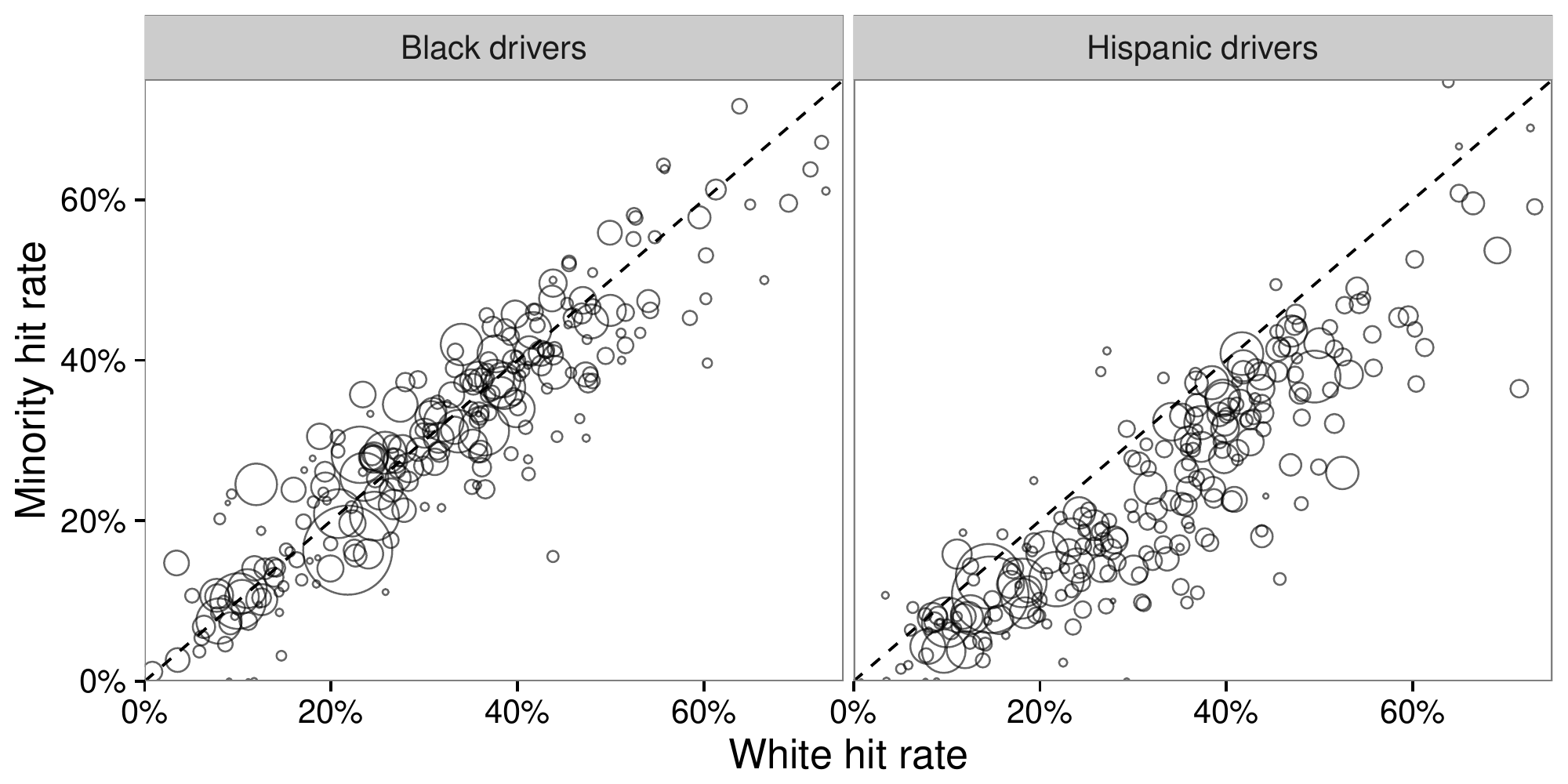}
\includegraphics[height=5cm]{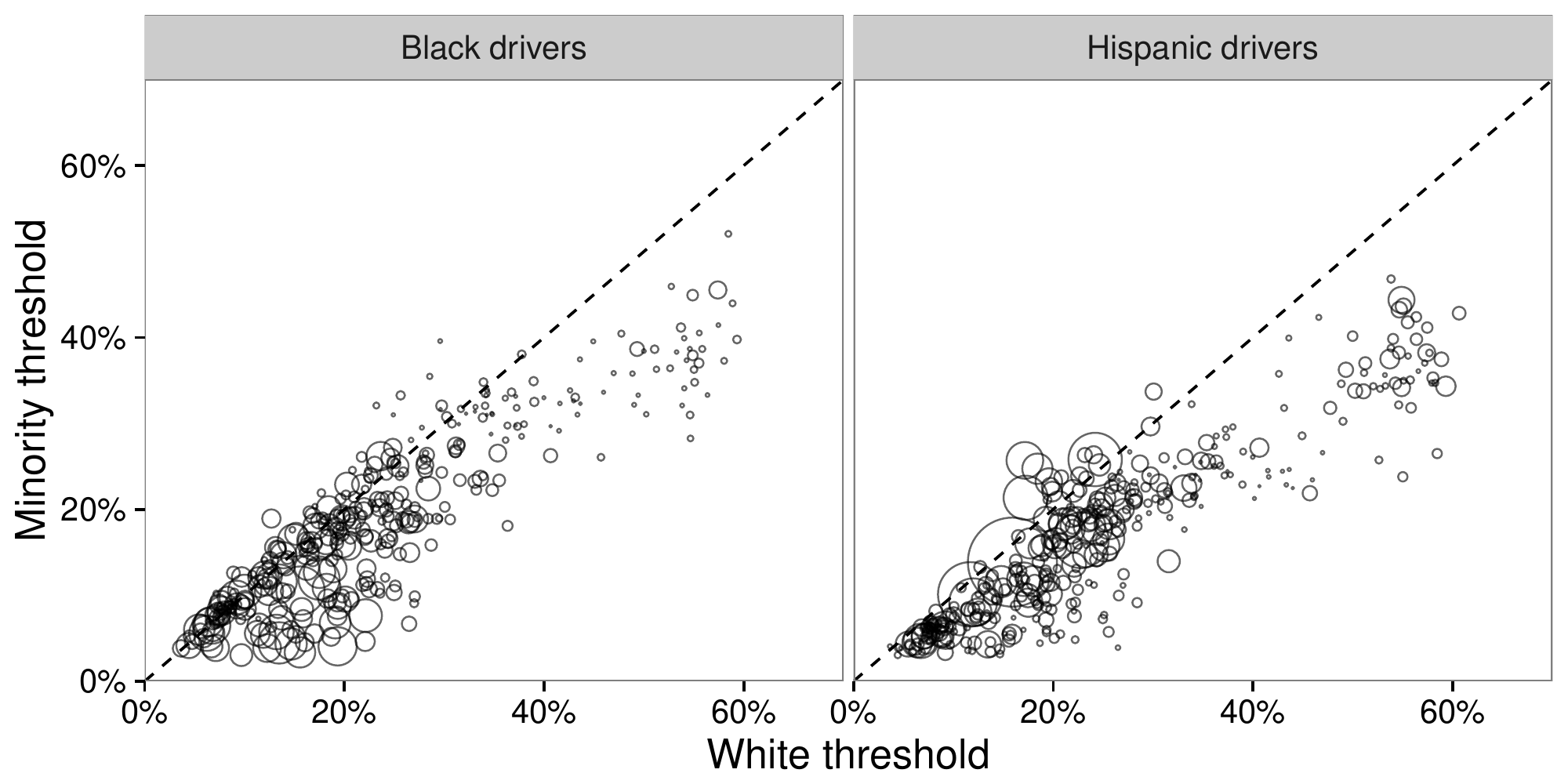}
\caption{\emph{Hit rates (top) and inferred search thresholds (bottom) by race and location,
covering 470,000 searches in 9 states.
Across locations, the inferred thresholds for
searching black and Hispanic drivers are typically lower than those for searching white drivers.
Despite these lower inferred search thresholds, 
hit rates for blacks are comparable to hit rates for whites, possibly due to the problem of \emph{infra-marginality} in outcome tests.
}}
\label{figure:hit_rate}
\end{figure}

In Figure~\ref{figure:hit_rate} (top row), we plot hit rates by race and location
for the nine states (Colorado, Connecticut, Illinois, North Carolina, Rhode Island, South Carolina, Texas, Washington, and Wisconsin) for which we have the necessary information: 
the race of the driver, the location of the stop, whether a search 
was conducted, and whether contraband was found.\footnote{
This information is also available for Vermont, but because of the
state's demographic composition, very few minorities are searched in any given county, and we thus exclude it from this analysis.}
Across jurisdictions, we consistently see that searches of Hispanic drivers are less successful than those of white drivers.
However, searches of white and black drivers generally have comparable hit rates. Aggregating across location, searches of Hispanic drivers yield contraband 22\% of the time, compared to 28\% for searches of white and black drivers. In computing these aggregate statistics,
we include Missouri and Maryland, which provide search and contraband data but not stop location, and Vermont, which has too few stops of minorities to be included in our county-level analysis in Figure~\ref{figure:hit_rate}.
The outcome test thus indicates that search decisions 
may be biased against Hispanic drivers but not 
black drivers.

\subsection{The threshold test}
\label{sec:threshold}
The outcome test is intuitively appealing, but it is not a perfect barometer of bias; in particular, it suffers from the problem of 
\emph{infra-marginality}~\citep{ayres2002,anwar2006}.
To illustrate this shortcoming, suppose that there are two, easily distinguishable types of white drivers: those who have a 5\% chance of carrying contraband, and those who have a 75\% chance of carrying contraband. Likewise assume that black drivers have either a 5\% or 50\% chance of carrying contraband. If officers search drivers who are at least 10\% likely to be carrying contraband, then searches of whites will be successful 75\% of the time whereas searches of blacks will be successful only 50\% of the time. Thus, although the search criterion is applied in a race-neutral manner, the hit rate for blacks is lower than the hit rate for whites, and the outcome test would (incorrectly) conclude searches are biased against black drivers.
The outcome test can similarly fail to detect discrimination when it is present.

To mitigate this limitation of outcome tests,
the \emph{threshold test} has been proposed as a more robust means for detecting discrimination~\citep{simoiu2016,pierson2017}.
This test aims to estimate race-specific probability thresholds above which officers search drivers---for example, the 10\% threshold in the hypothetical situation above.
Even if two race groups have the same observed hit rate, the threshold test may find that one group is searched on the basis of less evidence, indicative of discrimination.
To accomplish this task, 
the test simultaneously estimates race-specific 
search thresholds and risk distributions 
that are consistent with the observed search and hit rates across all jurisdictions. The threshold test can thus be seen as a hybrid between outcome and benchmark analysis.

Here we present a brief overview of the threshold test as applied in our setting; see \citet{simoiu2016} for a more complete description.
For each stop $i$, we assume that we observe:
(1) the race of the driver, $r_i$;
(2) the stop location, $d_i$;
(3) whether the stop resulted in a search, indicated by $S_i \in \{0,1\}$; and
(4) whether the stop resulted in a hit, indicated by $H_i \in \{0,1\}$.
We applied the threshold test separately on each state having the requisite data,
and limited to stop locations (e.g., counties) with at least 1,000 stops. 
If more than 100 locations in a state had over 1,000 stops, we considered only the 100 locations with the most stops.

The threshold test is based on a stylized model of officer behavior.
During each stop, officers observe a myriad of contextual factors---including the age and gender of the driver, the stop time and location, 
and behavioral indicators of nervousness or evasiveness.
We assume that officers distill these factors down to a single number that represents the likelihood the driver is carrying contraband, 
and then conduct a search if
that probability exceeds a fixed race- and location-specific threshold. 
Since there is uncertainty in who is pulled over in any given stop, the probability of finding contraband is modeled as a random draw from a race- and location-specific \emph{signal} distribution. 
The threshold test jointly estimates these search thresholds and signal distributions using
a hierarchical Bayesian model, as described below.
Under this model, lower search thresholds for one group relative to another are interpreted as evidence of \emph{taste-based} discrimination~\citep{becker1957}.

Formally, for each stop $i$, we assume $(S_i, H_i)$ is stochastically generated in three steps.
\begin{enumerate}
\item Given the race $r_i$ of the driver and the stop location $d_i$, the officer
observes a signal $p_i \sim \text{beta}(\phi_{r_id_i}, \lambda_{r_id_i})$,
where $\phi_{r_id_i}$ and $\lambda_{r_id_i}$ are defined by:
\begin{equation*}
\phi_{rd} = \textrm{logit}^{-1}(\phi_r + \phi_d)
\end{equation*}
and
\begin{equation*}
\lambda_{rd} = \exp(\lambda_r + \lambda_d).
\end{equation*}
The beta distribution is parameterized by its mean $\phi_{rd}$
and total count parameter $\lambda_{rd}$.
In terms of the standard count parameters $\alpha$ and $\beta$ of the beta distribution,
$\phi = \alpha/(\alpha + \beta)$ and $\lambda = \alpha + \beta$.
Thus, $\phi_{rd}$ is the overall probability that a stopped driver of race $r$ in location $d$ has contraband, 
and $\lambda_{rd}$ characterizes the heterogeneity of guilt across stopped drivers of that race in that location.
These parameters of the beta distributions
are in turn functions of parameters that depend separately on race and location.

\item $S_i = 1$ (i.e., a search is conducted) if and only if $p_i \geq t_{r_id_i}$.
The thresholds $t_{rd}$ are the key parameters of interest.

\item If $S_i = 1$, then $H_i \sim \text{Bernoulli}(p_i)$; otherwise $H_i = 0$.
\end{enumerate}

\noindent
This generative process is parameterized by
 $\{\phi_r\}$, $\{\lambda_r\}$, $\{\phi_d\}$, $\{\lambda_d\}$ and $\{t_{rd}\}$.
To complete the model specification, we place weakly informative priors on  $\phi_r$, $\lambda_r$, $\phi_d$, and $\lambda_d$, and 
place a weakly informative hierarchical prior on $t_{rd}$. 
The hierarchical structure allows us to make reasonable inferences even for  locations with a relatively small number of stops.
Finally, we compute the posterior distribution of the parameters given the data.

\begin{figure}[t]
\centering
\includegraphics[height=5cm]{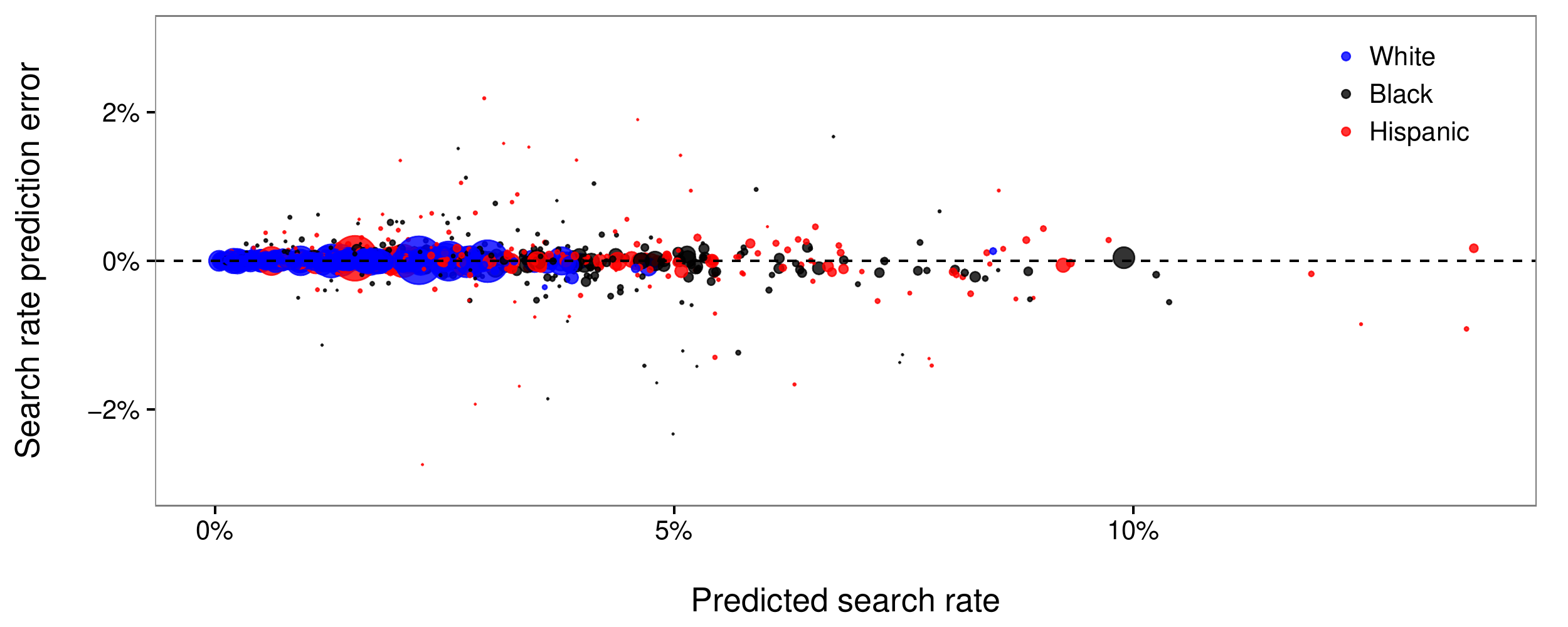}
\includegraphics[height=5cm]{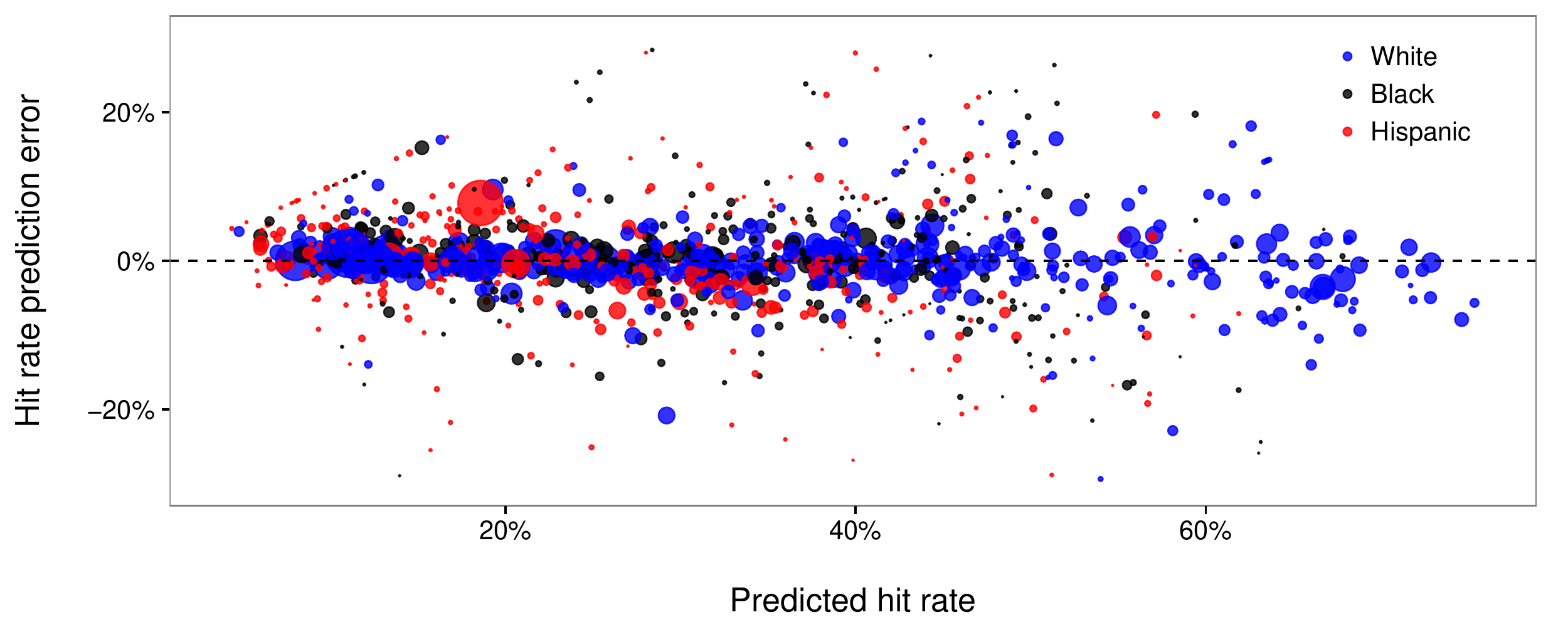}
\caption{\emph{Posterior predictive checks for search rates (top) and hit rates (bottom). Both plots indicate the fitted model captures key features of the data.
}}
\label{figure:ppc}
\end{figure}

We estimate the posterior distribution of the parameters
via Hamiltonian Monte Carlo (HMC) sampling~\citep{neal1994,duane1987}, a form of Markov chain Monte Carlo sampling~\citep{metropolis1953}.
We specifically use the No-U-Turn sampler (NUTS)~\citep{hoffman2013} as implemented in Stan~\citep{carpenter2016},
an open-source modeling language for full Bayesian statistical inference.
To assess convergence of the algorithm, we sampled five Markov chains in parallel and computed the potential scale reduction factor $\hat{R}$~\citep{gelman1992}. 
We found that 2,500 warmup iterations and 2,500 sampling iterations per chain were sufficient for convergence,
as indicated by $\hat{R}$ values less than $1.05$ for all parameters, as well as by visual inspection of the trace plots.

We apply \emph{posterior predictive checks}~\citep{gelman2014,gelman1996} to evaluate the extent to which the fitted model yields race- and location-specific search and hit rates that are in line with the observed data.
For each department and race group, we compare the observed search and hit rates to their expected values under the assumed data-generating process with parameters drawn from the inferred posterior distribution. Specifically, we compute the posterior predictive search and hit rates as follows. 
During model inference, our Markov chain Monte Carlo sampling procedure yields $2,500  \times 5 = 12,500$ draws from the
joint posterior distribution of the parameters. 
For each parameter draw---consisting of $\{\phi_r^*\}$, 
$\{\lambda_r^*\}$, $\{\phi_d^*\}$, $\{\lambda_d^*\}$ and $\{t_{rd}^*\}$---we analytically compute 
the search and hit rates $s_{rd}^*$ and $h_{rd}^*$ for each race-location pair implied by the data-generating process
with those parameters.
Finally, we average these search and hit rates over all 12,500 posterior draws.
Figure~\ref{figure:ppc} compares the model-predicted search and hit rates to the actual, observed values. 
Each point in the plot corresponds to a single race-location group, where groups are sized by number of stops.
The fitted model recovers the observed search rates almost perfectly across races and locations.
The fitted hit rates also agree with the data well,
with the largest groups exhibiting almost no error.
These posterior predictive checks thus indicate that the fitted model captures key features of the observed data.

We now turn to the substantive implications of our threshold analysis. 
As shown in Figure~\ref{figure:hit_rate} (bottom row), the threshold test indicates that
the bar for searching black and Hispanic drivers is 
lower than for searching white drivers in nearly every location we consider.
In aggregate, the inferred threshold for white drivers is 20\%, compared to 16\% for blacks and 14\% for Hispanics.
These aggregate thresholds are computed by taking a weighted average of location-specific thresholds, where weights are proportional to the total number of stops in each location.
The 95\% credible intervals for the aggregate, race-specific thresholds are non-overlapping:
$(19\%, 20\%)$ for white drivers,
$(15\%, 17\%)$ for black drivers,
and $(13\%, 14\%)$ for Hispanic drivers.
Whereas the outcome test indicates discrimination only against Hispanic drivers, 
the threshold test suggests discrimination against both blacks and Hispanics.
Consistent with past work~\citep{simoiu2016},
this difference appears to be driven by a small but disproportionate number of black drivers who have high inferred likelihood of carrying contraband. 
Thus, even though the threshold test finds the bar for searching black drivers is lower than for whites, these groups have similar hit rates.

The threshold test provides evidence of bias
in search decisions. 
However, as with all tests of discrimination, there is a limit to what one can conclude from such statistical analysis alone. 
For example, if search policies differ not only across but also within jurisdictions, then the threshold
test might mistakenly indicate discrimination where there is none. 
Additionally, if officers disproportionately suspect
more serious criminal activity when searching black and Hispanic drivers compared to whites, then
lower observed thresholds may stem from non-discriminatory police practices. 

\section{The effects of legalizing marijuana on stop outcomes}  

\begin{figure}[t]
\centering
\includegraphics[height=5cm]{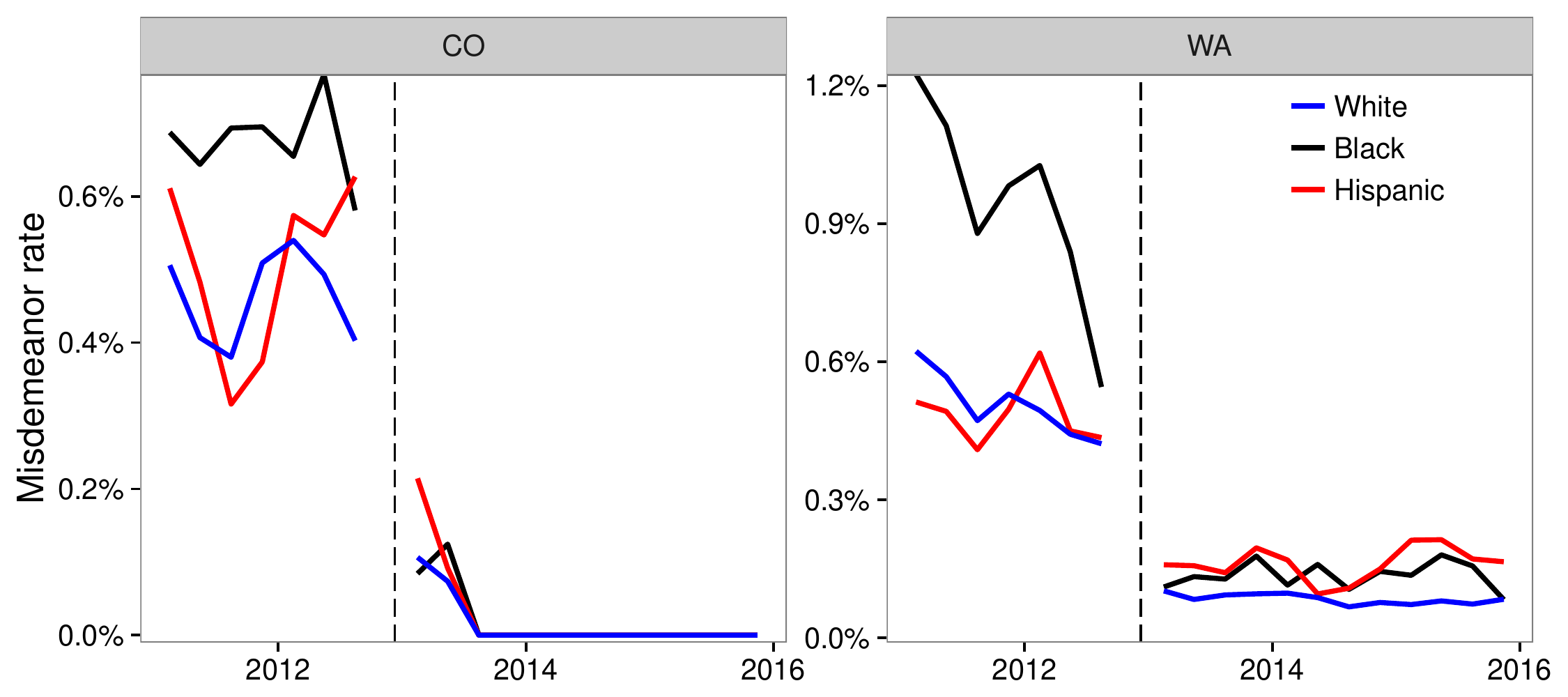}\\
\includegraphics[height=5cm]{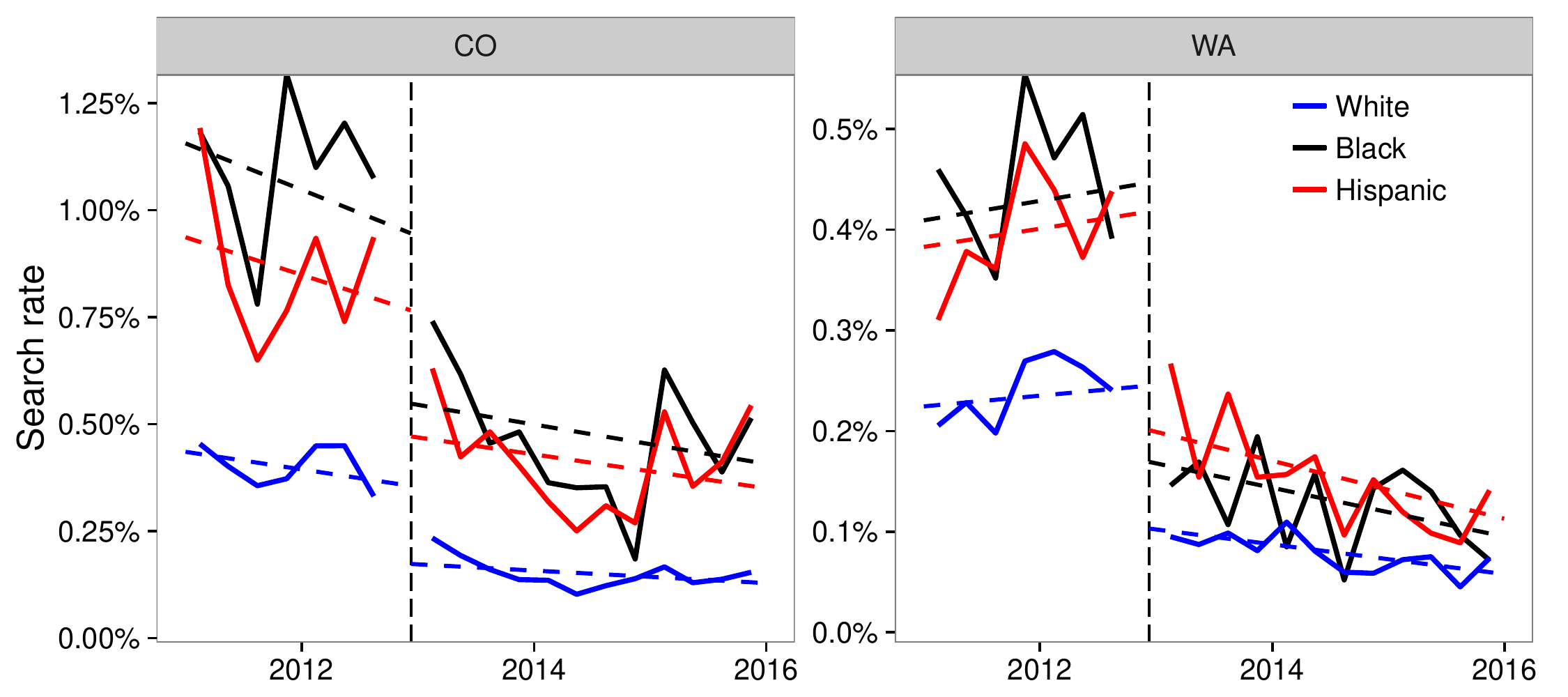}
\caption{\emph{The proportion of stops that result in a drug-related misdemeanor (top) or search (bottom) before and after recreational marijuana was legalized in Colorado and Washington at the end of 2012 (indicated by the vertical lines). 
Subsequent to legalization, there is a substantial drop in search and misdemeanor rates. The dashed lines show fitted linear trends pre- and post-legalization. 
}}
\label{figure:marijuana_search}
\end{figure}
 
\begin{figure}[t]
\centerline{
\includegraphics[height=9cm]{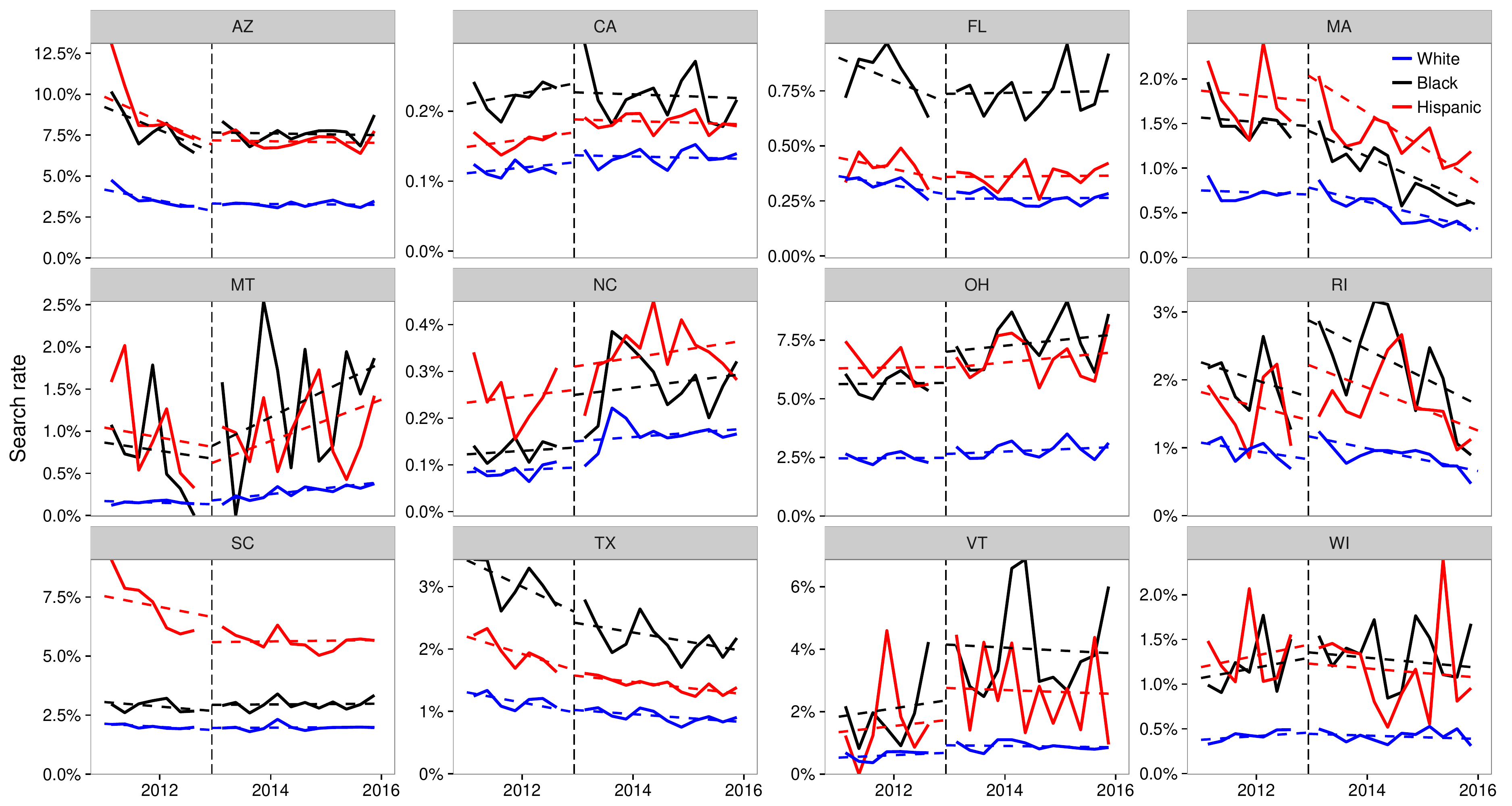}
}
\caption{\emph{
In the twelve states where marijuana was not legalized, and for which we have the necessary search data, 
search rates do not fall at the end of 2012; this pattern further suggests that marijuana legalization caused the observed drop in search rates in Colorado and Washington.
}}
\label{figure:marijuana_control}
\end{figure}

We conclude our analysis by 
investigating the effects of legalizing recreational marijuana on search and misdemeanor rates.
We specifically examine Colorado and Washington, the two states in which marijuana 
was recently legalized and for which we have detailed data.
As shown in Figure~\ref{figure:marijuana_search} (top) the number of drug-related misdemeanors in both states fell
substantially after marijuana was legalized at the end of 2012, in line with expectations.
In Colorado, we consider only misdemeanors for marijuana possession,
and so the rate necessarily drops after legalization; 
in Washington, we include misdemeanors
for any type of drug possession as more detailed information is not available, and so there are still some recorded drug misdemeanors post-legalization.
Notably, since black drivers were more likely to be charged with such 
offenses prior to legalization, 
black drivers were also disproportionately impacted by the policy change. 
This finding is consistent with past work showing
that marijuana laws disproportionately affect minorities~\citep{mitchell2015examining}.

Because the policy change decriminalized an entire class of behavior (i.e., possession of minor amounts of marijuana), it is not surprising that drug offenses correspondingly decreased. It is less clear, however, how the change affected officer behavior more broadly.
We find that after marijuana was legalized, the number of searches fell substantially in Colorado and Washington, (Figure~\ref{figure:marijuana_search}, bottom), 
ostensibly because the policy
change removed a common reason for conducting searches.
In both states, 
we exclude searches incident to an arrest and other searches that are conducted as a procedural matter, 
irrespective of any suspicion of drug possession.
Because black and Hispanic drivers were more likely to be searched prior to legalization,
the policy change reduced the absolute gap in search rates between white and minority drivers; however,
the relative gap persists, 
with minorities still more likely to be searched than whites.
We further note that marijuana legalization has secondary impacts for law-abiding drivers, as fewer searches overall means fewer searches of 
innocent individuals. In the year after legalization in Colorado and Washington, 40\% fewer drivers were searched with no contraband found than in the year before legalization. 

As shown in Figure~\ref{figure:marijuana_control}, in the twelve states where marijuana was not legalized---and for which we have the necessary search data---search rates did not drop significantly at the end of 2012.
This pattern further suggests that the observed drop in search rates in Colorado and Washington is due to marijuana legalization. 
To add quantitative detail to this visual result, we compute a simple difference-in-difference estimate~\citep{angrist2008}.
Specifically, we fit the following search model on
the set of stops in the 14 states we consider here
(Colorado, Washington, and the twelve non-legalization states in Figure~\ref{figure:marijuana_control}):
\begin{equation*}
\Pr(Y = 1) = 
\textrm{logit}^{-1}\left(
\sum_{s \in \text{state}}\beta_s I_s 
+ \sum_{r \in \text{race}}\beta_r I_r
+ \beta_t\cdot t
+ \sum_{r \in \text{race}}\alpha_r I_r Z
\right),
\end{equation*}
where $Y$ indicates whether a search was conducted,
$\beta_s$ and $\beta_r$ are state and race fixed effects,
and $\beta_t$ is a time trend, with $t$ a continuous variable in units of years since legalization (e.g., $t=0.5$ means 6 months post-legalization).
The $Z$ term indicates ``treatment'' status;
that is, $Z_i = 1$ in Colorado and Washington for stops carried out during the post-legalization period,
and $Z_i = 0$ otherwise.
Thus the key parameters of interest are the race-specific treatment effects $\alpha_r$.
Table~\ref{tab:weed} lists coefficients for the
fitted model.
We find that $\alpha_r$ is large and negative for whites, blacks, and Hispanics,
which again suggests the observed drop in searches in Colorado and Washington was due to the legalization of marijuana in those states.

\begin{table}[t]
\centering
\begin{tabular}{rrr}
 &  Coef. & s.e.\\ 
  \hline
  Effect of legalization on white drivers & -0.99 & 0.02 \\ 
  Effect of legalization on black drivers & -1.01 & 0.06 \\ 
  Effect of legalization on Hispanic drivers & -0.79 & 0.03\smallskip \\
  Time (years) & -0.02 & 0.00 \\ 
  Black driver & 0.79 & 0.00 \\ 
  Hispanic driver & 0.64 & 0.00 \\  
\end{tabular}
\caption{\emph{Effects of legalizing recreational marijuana on search rates, as estimated with a difference-in-difference model. All race groups experienced a large drop in search rate.}}
\label{tab:weed}
\end{table}

Despite marijuana legalization decreasing search rates for all races, 
Figure~\ref{figure:marijuana_search} shows that the relative disparity between whites and minorities remains. We adapt the threshold test to assess the extent to which this disparity in search rates may reflect bias.
Specifically, we estimate race-specific search thresholds pre- and post-legalization.
To do so, we first divide the stops into pre- and post-legalization periods, indexed by $t \in \{\textrm{pre},\,\textrm{post}\}$. The equations in Section \ref{sec:threshold} are modified to allow race-dependent time variation in the signal distributions and thresholds:
\begin{align*}
\phi_{rdt}&=\textrm{logit}^{-1}(\phi_r+\phi_d+\phi_{rt})\\
\lambda_{rdt}&=\exp(\lambda_r+\lambda_d+\lambda_{rt})\\
t_{rdt}&=t_{rd}+t_{rt},
\end{align*}
where the new parameters $\phi_{rt}$, $\lambda_{rt}$, and $t_{rt}$ are set to 0 when $t=\textrm{pre}$, and given a weakly informative $\text{N}(0,1)$ prior otherwise. 
Inference in the model is performed separately for Colorado and Washington.

\begin{figure}[t!]
\centerline{
\includegraphics[height=5cm]{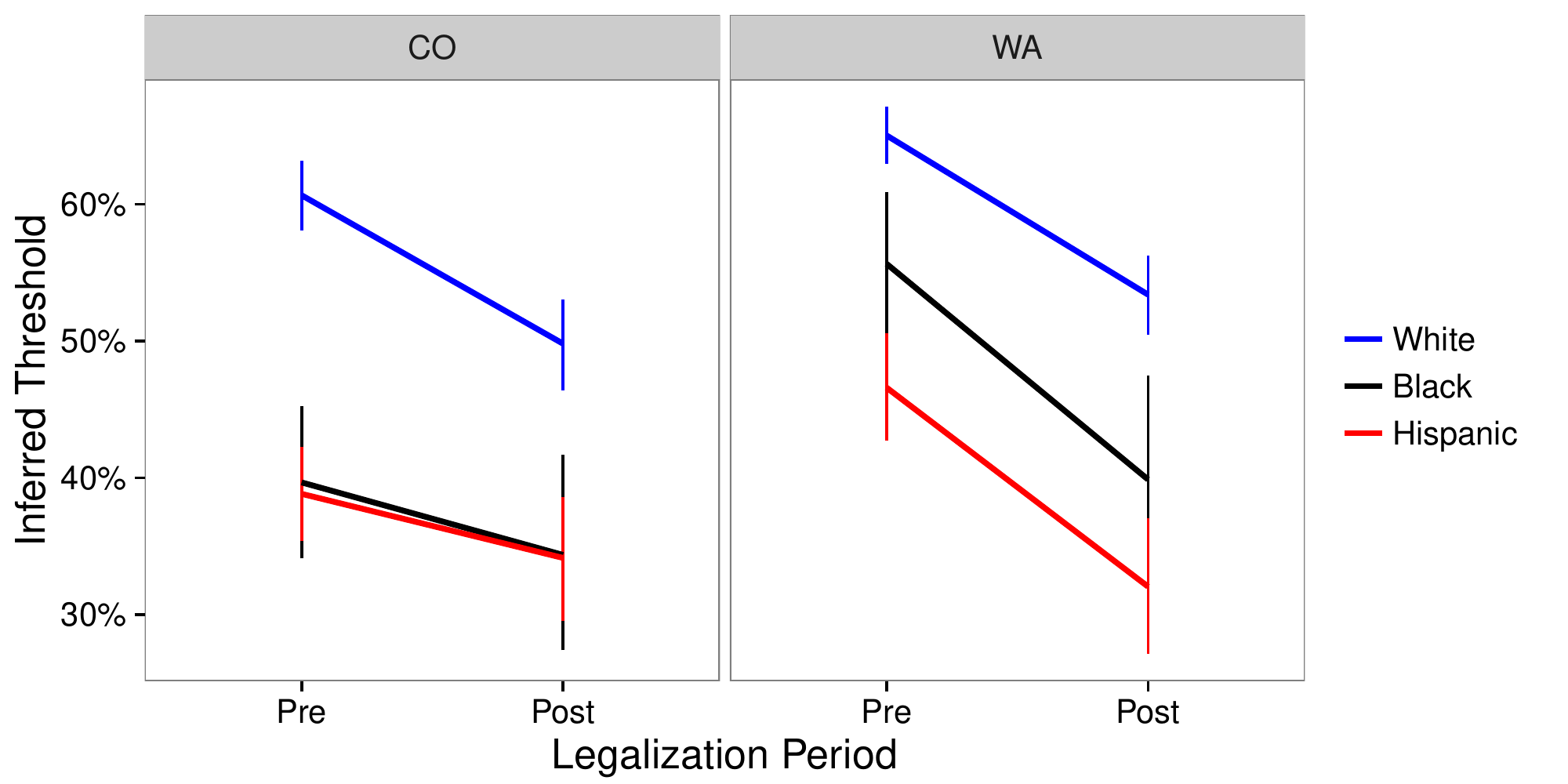}
}
\caption{\emph{
Inferred average thresholds faced by drivers of different races before and after marijuana legalization. Error bars show the 95\% credible intervals of the posterior thresholds. In all cases minority drivers face a lower threshold than white drivers.
}}
\label{figure:marijuana_threshold}
\end{figure}

Examining the inferred thresholds (shown in Figure~\ref{figure:marijuana_threshold}), we observe that whites drivers face consistently higher search thresholds than minority drivers, both before and after marijuana legalization. The data thus suggest that although overall search rates drop
in Washington and Colorado,
bias persists in search decisions.

Figure~\ref{figure:marijuana_threshold} also shows that the average threshold faced by all groups decreases after legalization (though not all drops are statistically significant). There are several possible explanations for this decrease. Officers may not have fully internalized the change of policy, searching people who would have been at risk of carrying contraband before legalization, but are no longer high risk now that marijuana is legal. Alternatively, or in addition, officers may now be focused on more serious offenses (such as drug trafficking), applying a lower threshold commensurate with the increase in the severity of the suspected crime. Finally, officers may have more resources after being relieved of the task of policing marijuana possession, freeing them to make searches with a lower chance of finding contraband.

\section*{Discussion}

Our investigation of over 60 million state patrol stops across the United States
reveals widespread racial disparities in stop, citation, search, and arrest rates.
It is important to note, however, that such differences may stem from a variety of mechanisms, and are not necessarily the result of racial bias. 
Moving beyond these disparities, a threshold analysis 
 indicates that black and Hispanic drivers are searched on the basis of less evidence than white drivers, suggestive of bias in search decisions.
The recent legalization of recreational marijuana in Colorado and Washington reduced the absolute gap in search rates between whites and minorities---because search rates decreased for all groups---but the relative gap remained. A threshold test further suggests that minorities continue to face bias in search decisions post-legalization.
In aggregate, our results lend insight into the differential impact of policing on minority communities nationwide.

Our study provides a unique perspective on working with large-scale policing data.
We conclude by offering several recommendations for 
data collection, release, and analysis.
At minimum, we encourage states to collect individual-level stop data that include 
the date and time of the stop;
the location of the stop;
the race, gender, and age of the driver; 
the stop reason;
whether a search was conducted;
the search type (e.g., ``probable cause'' or ``consent'');
whether contraband was found during a search;
the stop outcome (e.g., a citation or an arrest);
and the specific violation the driver was charged with.
Most states collect only a subset of this information.
There are also variables that are currently rarely collected 
but would be useful for analysis,
such as indicia of criminal behavior,
an officer's rationale for conducting a search,
and short narratives written by officers 
describing the incident.
New York City's UF-250 form for pedestrian stops is an example
of how such information can be efficiently collected~\citep{goel2016,mummolo2016}. 

Equally important to data collection is ensuring the 
integrity of the recorded information.
We frequently encountered missing values and errors in the data 
(e.g., implausible values for a driver's age and invalid racial categorizations).
Automated procedures can be put in place to help detect and correct such problems.
In most cases, the recorded race of the driver is based on the officer's perception,
rather than a driver's self-categorization.
While there are sound reasons for this practice, it increases the likelihood
of errors, 
a problem we observed in the Texas state patrol data.
To quantify and correct for this issue,
police departments might regularly audit their data,
possibly by comparing an officer's perception of race 
to a third party's judgment based on driver's license photos for
a random sample of stopped drivers.

Despite the existence of public records laws, 
seven states failed to respond to our repeated requests for information.
We hope law enforcement agencies consider taking steps to make data more accessible to external researchers and to the public.
Connecticut and North Carolina are at the forefront of opening up their data,
providing online portals for anyone to download and analyze this information. 

Finally, we hope that police departments start regularly analyzing
their data and report the results of their findings. 
Such analyses might include
estimates of stop, search, and hit rates, stratified by 
race, age, gender, and location; 
distribution of stop reasons by race; 
and trends over time. 
More ambitiously, departments could use 
their data to design statistically informed guidelines to encourage 
more consistent, efficient, and equitable decisions~\citep{goel2016,goel2016-risky}.
Many of these analyses can be automated and re-run regularly with little marginal effort.
In conjunction with releasing the data underlying these analyses,
we recommend the analysis code also be released to ensure reproducibility.
Collecting, releasing, and analyzing police data 
are essential steps for increasing the effectiveness and equity of law enforcement, 
and for improving relations with the public through transparency. 

\bibliographystyle{plainnat}
\bibliography{references}

\newpage
\section{Appendix}
\setcounter{table}{0}
\renewcommand{\thetable}{A\arabic{table}}

Below we describe the procedures we used to standardize each field in our data.

\begin{enumerate}
\item \textbf{Stop location.} The location of stops was encoded at varying levels of granularity across states, including police beat, city or town name, intersection, address, highway number and marker, county name, county FIPS code, district code, latitude and longitude coordinates, and highway patrol trooper zone. To provide a standard location coding, we aimed to map each stop to the county in which it occurred. For example, the data provided by Washington contained the highway type, number, and closest mile marker, which we first mapped to latitude and longitude coordinates using a publicly available dataset of highway marker locations; we then mapped the coordinates to counties using public shape files. Similarly, unidentified counties in Arizona were mapped to county using the highway the stop occurred on. For Connecticut, Massachusetts and Vermont, the counties were mapped using the police department that recorded the stop. For these states we found the county corresponding to the police department using the Google Maps API. In North Carolina, Illinois, and Rhode Island, no consistent county-level information was provided for state patrol stops. Therefore, we mapped stops to similarly granular location variables: for North Carolina we used district, for Illinois we used state patrol division, and for Rhode Island we used zone code. For North Carolina and Illinois we aggregated census statistics for the counties subsumed in the region to have a usable benchmark.

\item \textbf{Driver race.} We restrict our primary analysis to white, black, and Hispanic drivers. We specifically exclude stops of Asian and Native American drivers, as these groups were not sufficiently represented in our data to allow granular analysis. Some states provided ethnicity of the driver in addition to race; drivers with Hispanic ethnicity were considered Hispanic regardless of their recorded race, consistent with previous investigations. 
To aid future work, we classify drivers as ``Asian'' and ``other'' where possible---though these groups are not included in our main analysis. For example, Native American and Alaskan Native drivers were classified as ``other''; South Asian and Pacific Islander drivers were classified as ``Asian''.

\item \textbf{Driver age.} States provided either date of birth, birth year, or age of the driver. The age of the driver at the time of the stop was calculated by taking the difference between the stop date and the birth date of the driver, or stop year and birth year. If the inferred age of driver was less than 15 or greater than or equal to 100, we assumed the data were incorrect and treated age as missing in those cases. 

\item \textbf{Violation.} Some states listed one violation per stop, while others provided multiple violations (e.g., there were up to twelve recorded in Washington). If multiple violation codes were provided, all were included in our standardized data. The granularity of violation codes also varied greatly, from two categories (e.g., speeding and seat belt violations in Massachusetts) to over 1,000 in Colorado. Some states provided violation data by referring to local state statute numbers, which we mapped to a text description of the violation by consulting state traffic laws. We developed a two-level hierarchy of violation categories, and standardized each violation reason using this rubric. Our violation categories are as follows: (1) license/registration (with subcategories for license, registration/plates, and paperwork); 
(2) speeding; 
(3) seat belt violations; 
(4) stop sign/light; 
(5) equipment (with a subcategory for head/taillight violations); 
(6) driving under the influence (DUI);  
(7) moving violations (with subcategories for ``safe movement'' and ``cell phone''); 
and (8) truck violations. 
We coded violations using the most granular category possible. For states that had hundreds of violation codes, we mapped the most common ones until 95\% of stops were accounted for. 

\item \textbf{Stop purpose.} Some states distinguish between violation and stop purpose---the initial reason for the stop. When stop purpose was explicitly provided, it was placed in a separate column and normalized using the same values as the violation codes.

\item \textbf{Stop outcome.} Some states provided information on the outcome of the stop: for example, verbal warning, written warning, citation, summons, or arrest. In the case of speeding stops---which we specifically analyze---a stop was classified as a warning if either a warning or no penalty was given. A few states provided multiple outcomes for each stop, and in these cases, we recorded the most severe outcome---for example, if both a citation and a warning were given, the stop outcome was coded as a citation .

\item \textbf{Search conducted.} Many states provided a binary indicator for whether a search was conducted. In other cases we had to construct this field from other information in the data. For instance, North Carolina and South Carolina provided information on whether the driver, passenger, or vehicle was searched; we coded that a search was conducted if any of these three events occurred. 

\item \textbf{Search type.} We standardize search types into categories which include, for example, consent, probable cause, incident to arrest, inventory, warrant, protective frisk, and K9 searches.
Most of the standardization consisted of normalizing the language (e.g., ``drug dog alert'' and ``any K9 Used for Search'' were mapped to ``K9 search'').
Some states had multiple search reasons, others only one. If multiple search types were given, all were included in our standardized dataset.

\item \textbf{Contraband found.}
As with the ``search conducted'' field, states often provided a binary indicator for whether contraband was found. 
In other cases, it is constructed from multiple binary flags. For example, in South Carolina, we say that contraband was found if any of the ``Contraband'', ``ContrabandDrugs,'' ``ContrabandDrugParaphenalia,'', ``ContrabandWeapons'', or ``ContrabandDesc'' fields indicate that contraband was found. 
In some cases, it was indicated that contraband was found but no search was conducted. It is unclear whether a search was in fact conducted but not recorded, whether contraband was incorrectly marked,
or whether contraband was discovered through a process other than a search (e.g., found near the vehicle).
In these instances, we set the field value to ``false'', and note that the choice affects only a small proportion of searches and does not qualitatively affect our results.

\end{enumerate}

\begin{table}[t!]
\vspace{-1mm}
\centering
\small
\begin{tabular}{l c C{2cm} l}
\multirow{2}{*}{State} & Data released & Used in analysis & Response status\\
\hline
Alabama           &           &           &  No response                     \\ 
Alaska            &           &           &  Does not collect                \\ 
Arizona           & $\bullet$ & $\bullet$ &  Individual stop data received   \\ 
Arkansas          &           &           &  No central database             \\ 
California        & $\bullet$ & $\bullet$ &  Individual stop data received   \\ 
Colorado          & $\bullet$ & $\bullet$ &  Individual stop data received   \\ 
Connecticut       & $\bullet$ & $\bullet$ &  Individual stop data received   \\ 
Delaware          &           &           &  Provided reports only           \\ 
Florida           & $\bullet$ & $\bullet$ &  Individual stop data received   \\ 
Georgia           &           &           &  Does not collect                \\ 
Hawaii            &           &           &  No response                     \\ 
Idaho             &           &           &  Does not collect                \\ 
Illinois          & $\bullet$ & $\bullet$ &  Individual stop data received   \\ 
Indiana           &           &           &  No response                     \\ 
Iowa              & $\bullet$ &           &  Incomplete race data            \\ 
Kansas            &           &           &  Request denied                  \\ 
Kentucky          &           &           &  No central database             \\ 
Louisiana         &           &           &  Request denied                  \\ 
Maine             &           &           &  No central database             \\ 
Maryland          & $\bullet$ & $\bullet$ &  Individual stop data received   \\ 
Massachusetts     & $\bullet$ & $\bullet$ &  Individual stop data received   \\ 
Michigan          & $\bullet$ &           &  Incomplete race data            \\ 
Minnesota         &           &           &  Does not collect                \\ 
Mississippi       & $\bullet$ &           &  Incomplete race data            \\ 
Missouri          & $\bullet$ & $\bullet$ &  Summary data received           \\ 
Montana           & $\bullet$ & $\bullet$ &  Individual stop data received   \\
Nebraska          & $\bullet$ & $\bullet$ &  Summary data received           \\
Nevada            & $\bullet$ &           &  Incomplete race data            \\
New Hampshire     & $\bullet$ &           &  Incomplete race data            \\
New Jersey        & $\bullet$ & $\bullet$ &  Individual stop data received   \\
New Mexico        &           &           &  No response                     \\
New York          &           &           &  No central database        \\
North Carolina    & $\bullet$ & $\bullet$ &  Individual stop data received   \\
North Dakota      & $\bullet$ &           &  Provided citation data only     \\
Ohio              & $\bullet$ & $\bullet$ &  Individual stop data received   \\
Oklahoma          &           &           &  No response                     \\
Oregon            & $\bullet$ &           &  Summary data received, not usable\\
Pennsylvania      &           &           &  Request denied                  \\
Rhode Island      & $\bullet$ & $\bullet$ &  Individual stop data received   \\
South Carolina    & $\bullet$ & $\bullet$ &  Individual stop data received   \\
South Dakota      & $\bullet$ &           &  Missing race data               \\
Tennessee         & $\bullet$ &           &  Provided citation data only     \\
Texas             & $\bullet$ & $\bullet$ &  Individual stop data received   \\
Utah              &           &           &  Request denied                  \\
Vermont           & $\bullet$ & $\bullet$ &  Individual stop data received   \\
Virginia          & $\bullet$ &           &  Summary data received, not usable   \\
Washington        & $\bullet$ & $\bullet$ &  Individual stop data received   \\
West Virginia     &           &           &  No central database             \\
Wisconsin         & $\bullet$ & $\bullet$ &  Individual stop data received   \\
Wyoming           & $\bullet$ &           &  Provided citation data only     \\  
\end{tabular}
\caption{\emph{Status of responses to our public record requests, 
at time of writing.}} 
\label{tab:states_status}
\end{table}

\afterpage{
\begin{landscape}
\thispagestyle{empty}
\begin{table}[ht!]
\centerline{
\small
\begin{tabular}{clrlcccccccccccccc}
 &  \multirow{2}{*}{State}  & \thead{\multirow{2}{*}{Stops}} & \thead{Time}  & \thead{Stop} & \thead{Stop} & \thead{Stop} & \thead{Driver} & \thead{Driver} & \thead{Driver} & \thead{Stop} & \thead{Search} & \thead{Search} & \thead{Contraband} & \thead{Stop}  \\
 &                          &                                & \thead{Range} & \thead{Date} & \thead{Time} & \thead{Location} & \thead{Race} & \thead{Gender} & \thead{Age} & \thead{Reason} & \thead{Conducted} & \thead{Type} & \thead{Found} & \thead{Outcome}  \\
  \hline
1 & \textbf{Arizona} &   2,251,992 & 2009-2015 & $\bullet$ & $\bullet$ & $\bullet$ & $\bullet$ & $\bullet$ &  &  & $\bullet$ &  & $\bullet$ & $\bullet$ \\ 
  2 & \textbf{California} &  31,778,515 & 2009-2016 & $\bullet$ &  & $\bullet$ & $\bullet$ & $\bullet$ &  & $\bullet$ & $\bullet$ & $\bullet$ &  & $\bullet$ \\ 
  3 & \textbf{Colorado} &   2,584,744 & 2010-2016 & $\bullet$ & $\bullet$ & $\bullet$ & $\bullet$ & $\bullet$ & $\bullet$ & $\bullet$ & $\bullet$ & $\bullet$ & $\bullet$ &  \\ 
  4 & \textbf{Connecticut} &     318,669 & 2013-2015 & $\bullet$ & $\bullet$ & $\bullet$ & $\bullet$ & $\bullet$ & $\bullet$ & $\bullet$ & $\bullet$ & $\bullet$ & $\bullet$ & $\bullet$ \\ 
  5 & \textbf{Florida} &   5,421,446 & 2010-2016 & $\bullet$ & $\bullet$ & $\bullet$ & $\bullet$ & $\bullet$ & $\bullet$ & $\bullet$ & $\bullet$ & $\bullet$ &  & $\bullet$ \\ 
  6 & \textbf{Illinois} &   4,715,031 & 2004-2015 & $\bullet$ & $\bullet$ & $\bullet$ & $\bullet$ & $\bullet$ & $\bullet$ & $\bullet$ & $\bullet$ & $\bullet$ & $\bullet$ & $\bullet$ \\ 
  7 & Iowa &   2,441,335 & 2006-2016 & $\bullet$ & $\bullet$ &  &  &  &  & $\bullet$ &  &  &  & $\bullet$ \\ 
  8 & \textbf{Maryland} &   1,113,929 & 2007-2014 &  &  &  & $\bullet$ & $\bullet$ &  & $\bullet$ & $\bullet$ &  & $\bullet$ & $\bullet$ \\ 
  9 & \textbf{Massachusetts} &   3,418,298 & 2005-2015 & $\bullet$ &  & $\bullet$ & $\bullet$ & $\bullet$ & $\bullet$ &  & $\bullet$ & $\bullet$ & $\bullet$ & $\bullet$ \\ 
  10 & Michigan &     709,699 & 2001-2016 & $\bullet$ & $\bullet$ & $\bullet$ &  &  &  & $\bullet$ &  &  &  & $\bullet$ \\ 
  11 & Mississippi &     215,304 & 2013-2016 & $\bullet$ &  & $\bullet$ & $\bullet$ & $\bullet$ & $\bullet$ & $\bullet$ &  &  &  &  \\ 
  12 & \textbf{Missouri} &   2,292,492 & 2010-2015 &  &  &  & $\bullet$ &  &  &  & $\bullet$ &  & $\bullet$ &  \\ 
  13 & \textbf{Montana} &     825,118 & 2009-2016 & $\bullet$ & $\bullet$ & $\bullet$ & $\bullet$ & $\bullet$ & $\bullet$ & $\bullet$ & $\bullet$ & $\bullet$ &  & $\bullet$ \\ 
  14 & \textbf{Nebraska} &   4,277,921 & 2002-2014 &  &  &  & $\bullet$ &  &  &  & $\bullet$ &  &  &  \\ 
  15 & Nevada &     737,294 & 2012-2016 & $\bullet$ &  &  & $\bullet$ &  & $\bullet$ & $\bullet$ &  &  &  & $\bullet$ \\ 
  16 & New Hampshire &     259,822 & 2014-2015 & $\bullet$ & $\bullet$ & $\bullet$ &  & $\bullet$ &  & $\bullet$ &  &  &  & $\bullet$ \\ 
  17 & \textbf{New Jersey} &   3,845,335 & 2009-2016 & $\bullet$ & $\bullet$ & $\bullet$ & $\bullet$ & $\bullet$ &  & $\bullet$ &  &  &  & $\bullet$ \\ 
  18 & \textbf{North Carolina} &   9,558,084 & 2000-2015 & $\bullet$ &  & $\bullet$ & $\bullet$ & $\bullet$ & $\bullet$ & $\bullet$ & $\bullet$ & $\bullet$ & $\bullet$ & $\bullet$ \\ 
  19 & North Dakota &     330,063 & 2010-2015 & $\bullet$ & $\bullet$ & $\bullet$ & $\bullet$ & $\bullet$ & $\bullet$ & $\bullet$ &  &  &  &  \\ 
  20 & \textbf{Ohio} &   6,165,997 & 2010-2015 & $\bullet$ & $\bullet$ & $\bullet$ & $\bullet$ & $\bullet$ &  &  & $\bullet$ &  &  &  \\ 
  21 & Oregon  &   1,143,017 & 2010-2016 &  &  &  & $\bullet$ &  &  &  &  &  &  &  \\ 
  22 & \textbf{Rhode Island} &     509,681 & 2005-2015 & $\bullet$ & $\bullet$ & $\bullet$ & $\bullet$ & $\bullet$ & $\bullet$ & $\bullet$ & $\bullet$ & $\bullet$ & $\bullet$ & $\bullet$ \\ 
  23 & \textbf{South Carolina} &   8,440,934 & 2005-2016 & $\bullet$ &  & $\bullet$ & $\bullet$ & $\bullet$ & $\bullet$ &  & $\bullet$ &  & $\bullet$ & $\bullet$ \\ 
  24 & South Dakota &     281,249 & 2012-2015 & $\bullet$ & $\bullet$ & $\bullet$ &  & $\bullet$ &  & $\bullet$ &  &  &  & $\bullet$ \\ 
  25 & Tennessee &   3,829,082 & 1996-2016 & $\bullet$ & $\bullet$ & $\bullet$ & $\bullet$ & $\bullet$ &  & $\bullet$ &  &  &  & $\bullet$ \\ 
  26 & \textbf{Texas} &  23,397,249 & 2006-2015 & $\bullet$ & $\bullet$ & $\bullet$ & $\bullet$ & $\bullet$ &  & $\bullet$ & $\bullet$ & $\bullet$ & $\bullet$ & $\bullet$ \\ 
  27 & \textbf{Vermont} &     283,285 & 2010-2015 & $\bullet$ & $\bullet$ & $\bullet$ & $\bullet$ & $\bullet$ & $\bullet$ & $\bullet$ & $\bullet$ & $\bullet$ & $\bullet$ & $\bullet$ \\ 
  28 & Virginia  &   5,006,725 & 2006-2016 & $\bullet$ &  & $\bullet$ & $\bullet$ &  &  &  & $\bullet$ &  &  &  \\ 
  29 & \textbf{Washington} &   8,624,032 & 2009-2016 & $\bullet$ & $\bullet$ & $\bullet$ & $\bullet$ & $\bullet$ & $\bullet$ & $\bullet$ & $\bullet$ & $\bullet$ & $\bullet$ & $\bullet$ \\ 
  30 & \textbf{Wisconsin} &   1,059,033 & 2010-2016 & $\bullet$ & $\bullet$ & $\bullet$ & $\bullet$ & $\bullet$ &  & $\bullet$ & $\bullet$ & $\bullet$ & $\bullet$ & $\bullet$ \\ 
  31 & Wyoming  &     173,455 & 2011-2012 & $\bullet$ & $\bullet$ & $\bullet$ & $\bullet$ & $\bullet$ & $\bullet$ & $\bullet$ &  &  &  &  \\ 
   & \textbf{Total} & \textbf{136,008,830} &  &  &  &  &  &  &  &  &  &  &  &  \\ 
\end{tabular}
}
\vspace{2mm}
\caption{\emph{Overview of the complete state patrol dataset. For each column a solid circle signifies data are available for at least 70\% of the stops. The states used for the analysis in the paper are boldfaced. For all states except Illinois, North Carolina, and Rhode Island, ``stop location'' refers to county; for these three states, it refers to a similarly granular location variable, as described above.}} 
\label{tab:summary_complete}
\end{table}
\end{landscape}
}

\afterpage{
\begin{landscape}
\thispagestyle{empty}
\begin{table}[ht]
\centerline{
\small
\begin{tabular}{lllll}
Outcome & Covariates & Black & Hispanic & States \\ 
  \hline
  Stop (negative binomial) & race, location, demo, year & 0.37 (0.01) & -0.40 (0.01) & CO,CT,FL,IL,MA,MT,NC,SC,VT,WA \\ 
  Stop (Poisson) & race, location, demo, year & 0.27 (0.01) & -0.27 (0.01) & CO,CT,FL,IL,MA,MT,NC,SC,VT,WA \\ 
  Stop (quasi-Poisson) & race, location, demo, year & 0.27 (0.00) & -0.27 (0.01) & CO,CT,FL,IL,MA,MT,NC,SC,VT,WA \\ 
  \hline
  Citation & race & 0.61 (0.00) & -0.11 (0.00) & CO,FL,IL,MT,NC,RI,TX,WI \\ 
  Citation & race, location & 0.22 (0.00) & 0.36 (0.00) & CO,FL,IL,MT,NC,RI,TX,WI \\ 
  Citation & race, location, time & 0.24 (0.00) & 0.36 (0.00) & CO,FL,IL,MT,RI,TX,WI \\ 
  Citation & race, location, demo & 0.17 (0.00) & 0.32 (0.00) & CO,FL,IL,MT,NC,RI \\ 
  Citation & race, location, time, demo & 0.18 (0.00) & 0.29 (0.00) & CO,FL,IL,MT,RI \\ 
  \hline
  Search & race & 0.57 (0.00) & 0.64 (0.00) & AZ,CA,CO,CT,FL,IL,MA,MD,MO,MT,NC,NE,OH,RI,SC,TX,VT,WA,WI \\ 
  Search & race, location & 0.68 (0.00) & 0.74 (0.00) & AZ,CA,CO,CT,FL,IL,MA,MT,NC,OH,RI,SC,TX,VT,WA,WI \\ 
  Search & race, location, time & 0.75 (0.00) & 0.67 (0.00) & AZ,CO,CT,FL,IL,MT,OH,RI,TX,VT,WA,WI \\ 
  Search & race, location, demo & 0.56 (0.00) & 0.66 (0.01) & CO,CT,FL,IL,MA,MT,NC,RI,SC,VT,WA \\ 
  Search & race, location, time, demo & 0.73 (0.01) & 0.54 (0.01) & CO,CT,FL,IL,MT,RI,VT,WA \\ 
  \hline
  Consent search & race & 0.66 (0.01) & 1.14 (0.01) & CO,FL,MA,MD,NC,TX,WA \\ 
  Consent search & race, location & 0.76 (0.01) & 0.76 (0.01) & CO,FL,MA,NC,TX,WA \\ 
  Consent search & race, location, time & 0.77 (0.01) & 0.76 (0.01) & CO,FL,TX,WA \\ 
  Consent search & race, location, demo & 0.69 (0.02) & 0.70 (0.02) & CO,FL,MA,NC,WA \\ 
  Consent search & race, location, time, demo & 0.77 (0.03) & 0.62 (0.02) & CO,FL,WA \\ 
  \hline
  Arrest & race & 0.51 (0.00) & 0.72 (0.00) & AZ,CA,CO,CT,FL,MA,MD,MT,NC,OH,RI,SC,VT,WI \\ 
  Arrest & race, location & 0.50 (0.00) & 0.61 (0.00) & AZ,CA,CO,CT,FL,MA,MT,NC,OH,RI,SC,VT,WI \\ 
  Arrest & race, location, time & 0.75 (0.01) & 0.75 (0.01) & AZ,CO,CT,FL,MT,OH,RI,VT,WI \\ 
  Arrest & race, location, demo & 0.44 (0.00) & 0.75 (0.01) & CO,CT,FL,MA,MT,NC,RI,SC,VT \\ 
  Arrest & race, location, time, demo & 0.65 (0.01) & 0.69 (0.01) & CO,CT,FL,MT,RI,VT \\ 
\end{tabular}
}
\caption{\emph{Coefficients for driver race for various regression specifications, with
standard errors in parentheses. 
``Covariates'' denotes the set of variables used in the regression; 
``time'' indicates that stop year, stop quarter, stop weekday, and stop hour are included as covariates, 
and ``demo'' indicates that driver age and gender are included as covariates. 
Logistic regression is used in all cases except for estimating stop rate.}}
\label{tab:all_regression_coefficients}
\end{table}
\end{landscape}
}

\end{document}